\documentclass[aps,pre,nofootinbib,amsfonts,superscriptaddress,longbibliography,showkeys,notitlepage]{revtex4-1}
\usepackage[utf8]{inputenc}
\usepackage[T1]{fontenc}
\usepackage[english]{babel}
\usepackage{graphicx,color}

\usepackage{booktabs} 

\usepackage{xcolor}
\usepackage{hyperref}
\hypersetup{
    colorlinks,
    linkcolor={red!80!black},
    citecolor={blue!50!black},
    urlcolor={blue!80!black}
}

\usepackage{amsmath}
\usepackage{amssymb}
\usepackage{bm}

\usepackage{subfigure}
\usepackage{color}

\usepackage{tabularx}
\usepackage{booktabs}   
\newcommand{\ra}[1]{\renewcommand{\arraystretch}{#1}}

\newcolumntype{R}[1]{>{\raggedleft\let\newline\\\arraybackslash\hspace{0pt}}m{#1}}
\usepackage{threeparttable}
\usepackage{colortbl}
\usepackage{multirow}

\usepackage[super]{nth}

\DeclareMathOperator{\area}{area}

\DeclareMathOperator{\closest}{closest}
\DeclareMathOperator{\distance}{distance}

\DeclareMathOperator{\wdist}{wdist}

\makeatletter
\newcommand*{\rom}[1]{\expandafter\@slowromancap\romannumeral #1@}
\makeatother

\usepackage{cleveref}
\newcommand{\bst}[1]{\textcolor[RGB]{27,158,119}{\mathbf{#1}}}

\usepackage{afterpage}

\begin{document}

\title{Socio-economic, built environment, and mobility conditions associated with crime: A study of multiple cities}

\author{Marco De Nadai}
\affiliation{University of Trento, Trento, Italy}
\affiliation{Fondazione Bruno Kessler, Trento, Italy}

\author{Yanyan Xu}
\affiliation{Massachusetts Institute of Technology Cambridge, USA}

\author{Emmanuel Letouzé}
\affiliation{Data-pop Alliance, New York, USA}

\author{Marta C. González}
\affiliation{University of California Berkeley, Berkeley, USA}

\author{Bruno Lepri}
\affiliation{Fondazione Bruno Kessler, Trento, Italy}

\begin{abstract}
Nowadays, 23\% of the world population lives in multi-million cities. In these metropolises, criminal activity is much higher and violent than in either small cities or rural areas.
Thus, understanding what factors influence urban crime in big cities is a pressing need. Mainstream studies analyse crime records through historical panel data or analysis of historical patterns combined with ecological factor and exploratory mapping. More recently, machine learning methods have provided informed crime prediction over time. However, previous studies have focused on a single city at a time, considering only a limited number of factors (such as socio-economical characteristics) and often at large spatial units. Hence, our understanding of the factors influencing crime across cultures and cities is very limited. 
Here we propose a Bayesian model to explore how crime is related not only to socio-economic factors but also to the built environmental (e.g. land use) and mobility characteristics of neighbourhoods. 
To that end, we integrate multiple open data sources with mobile phone traces and compare how the different factors correlate with crime in diverse cities, namely Boston, Bogotá, Los Angeles and Chicago. We find that the combined use of socio-economic conditions, mobility information and physical characteristics of the neighbourhood effectively explain the emergence of crime, and improve the performance of the traditional approaches.
However, we show that the socio-ecological factors of neighbourhoods relate to crime very differently from one city to another. Thus there is clearly no "one fits all" model.
\end{abstract}

\flushbottom
\maketitle
\thispagestyle{empty}

\section*{Introduction}
In criminology, social cohesion among neighbours has been linked to their willingness to cooperate in order to solve common problems and reduce violence~\cite{Graif2009, Sampson1989, Sampson1997}. Cooperation, as opposed to disorganization of neighbours is indeed believed to create the mechanisms by which residents themselves achieve guardianship and public order~\cite{Sampson1989}. 
This mechanism also finds its roots in urban planning, where the relationship between specific aspects of urban architecture~\cite{newman1972defensible} and urban physical characteristics~\cite{jacobs1961death} are related to security.
However, neighbourhoods are not to be considered islands unto themselves, as they are embedded in a city-wide system of social interactions. On a daily basis, people's routine exposes residents to different conditions, possibilities~\cite{wang2018urban}, and it may favour crime~\cite{cohen1979social}.
Yet, mainstream studies focus on just a subset of static factors at a time, often in a single city (e.g. Chicago or New York), thus neglecting the complex urban interplay between crime, people, places, culture and human mobility.

Criminology widely recognize the importance of places. Crime occurs in small areas such as street segments, buildings or parks. However, neighbourhoods and their contextual characteristics are also believed to influence offenders' activities. Studies on small areas and neighbourhoods roughly come from two streams of literature.
The first stream focuses on the routine activity and crime pattern theories~\cite{cohen1979social, felson1998opportunity, brantingham1993nodes}, and small areas.
These studies suggest that crime occurs when an offender, its suitable target, and the absence of any deterrence system, such as police or even ordinary citizens~\cite{felson2010crime}, converge at a place. 
The presence of people influence the number of offenders and targets, but daily routine of residents exposes homes and people to predatory crimes~\cite{hindelang1978victims}.
The built environment was also found to affect criminal activities, as physical disorder and specific locations (e.g. bar, taverns) attract offenders and suitable targets~\cite{o2015public, murray2008measuring, salesses2013collaborative}.
The second stream of literature builds upon the social disorganization theory~\cite{Sampson1989, Sampson1997}, which found high crime concentration in socially and economically disadvantaged neighbourhoods. 
In these studies, census data is the primary source used to measure social cohesion through socio-economic disadvantage, ethnic diversity, residential instability~\cite{Sampson1989, Sampson1997, doi:10.1177/0022427885022001002}. 
In some cases, new sources of data were used. For example, scholars exploited synthetic social ties to simulate neighbourhood cohesion~\cite{hipp2013extrapolative}, and mobility flows to indicate crime opportunities and connections between neighbourhoods~\cite{song2019crime}. Others leveraged crowd-sourced Point of Interests (POIs), taxi flows~\cite{Wang:2016:CRI:2939672.2939736}, and dynamic population mapping from satellite imagery~\cite{andresen2006crime, andresen2011ambient} and mobile phone activity~\cite{bogomolov2014once, malleson2015spatio} to assess the presence of people. 
Altogether, these results highlight the tight relation between socio-economic, built environment and mobility conditions, and their impact on criminal activities.
Although the two streams of theory are often seen as competing, we argue that they can complement each other.
However, very limited work has integrated socio-economic, built environment and mobility conditions together in multiple cities and in small areas. Existing literature focuses on  a single city, and often describe crimes at the neighbourhood level and rely on census boundaries.
These limitations result in a fragmented and incomplete picture of how the numerous factors influence crime in the urban context and limit the impact of the conclusions.

Here, we seek to shed light on the diverse set of factors at play with urban crime exploring how this is related, at the same time, to social disorganisation, built environment characteristics and human mobility. Specifically, we analyse crime at the level of blocks, considering both the local features of the block and its surrounding context, represented by all the blocks within a half-mile.
The contribution of this paper is twofold. 
First, we address the need for a comprehensive study that explores crime patterns at fine grained resolution across multiple cities of the world, analysing Bogotá, Boston, Los Angeles and Chicago. 
Secondly, we show that the previously neglected complex interplay between crime, people, places, and human mobility can significantly improve the performance of the crime inference. We make use of massive and ubiquitous data sources such as mobile phone records and geographical data, implying that the resulting framework can be replicated at scale.
Our generated insights can help recommend effective policies and interventions that improve urban security.

\section*{Results}
We study criminal activity in Bogotá (Colombia), Boston (USA), Chicago (USA) and Los Angeles (USA), four very different cities with respect to cultural, urban and socio-economic conditions.
The selected unit of analysis is the census block group, the smallest geographical unit for which the census publishes data, and measuring on average 378 square meters.
We account for the contextual characteristics around the block group, here called \emph{core}, by computing a \emph{corehood}, defined as the set of all the surrounding block groups within a half mile from the core (see \Cref{fig:eff}). Note that neighbouring cores have overlapping corehoods.
We tested different sizes of the corehood, finding the half mile distance as the best to describe the neighborhood effect (see the Supplementary Information (SI) Note 11).

\begin{figure}[t!]
	\centering
	\includegraphics[width=0.8\columnwidth]{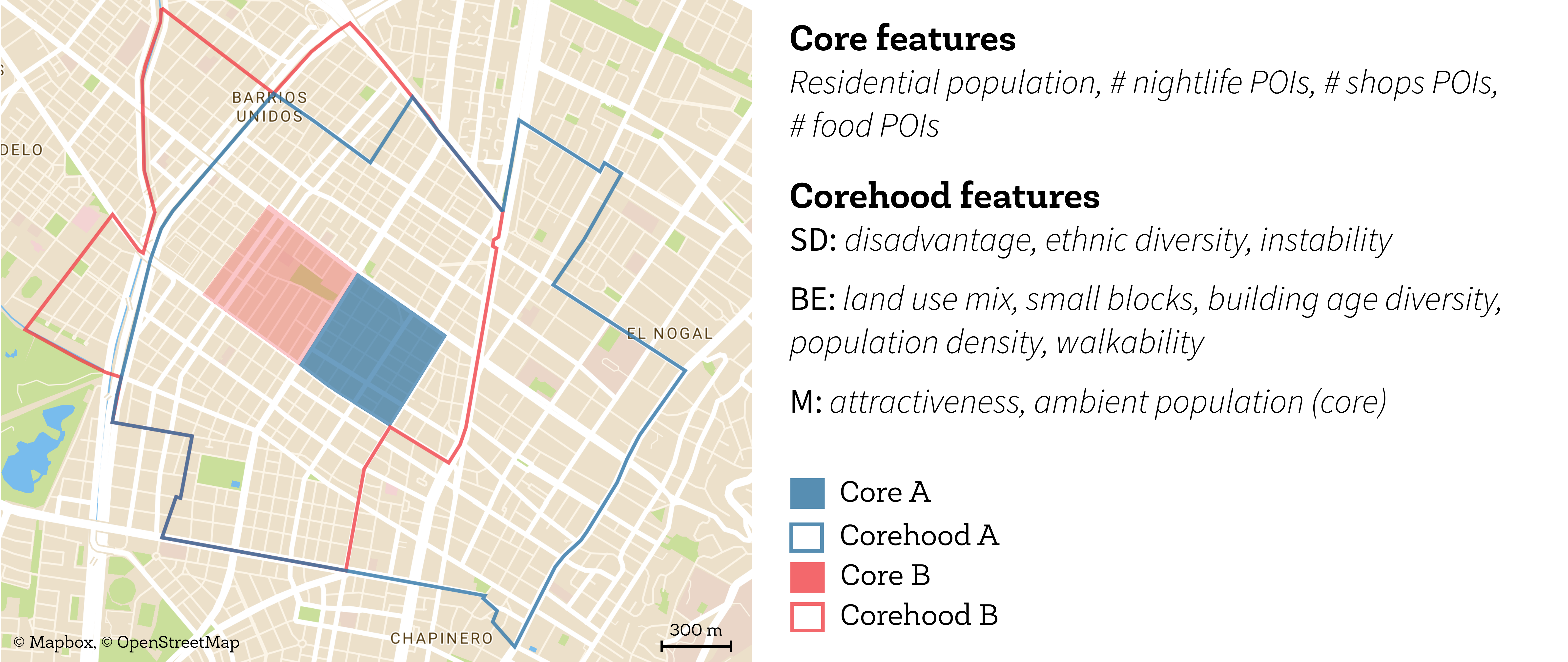}
	\caption{For each block group (the core), we consider the block groups within a half mile as its corehood. Blocks that are near each share most of their corehood. In this example, we show two cores in Bogotá and their corresponding corehood. We focus on three aspects of the core and the corehood: the Social Disorganization (SD), the Built Environment (BE), and the Mobility (M). The core, where crime is predicted, measures on average 378 square meters.} 
	\label{fig:eff}
\end{figure}

Criminal activity is provided by police agencies, which record through police reports the geographic location, date, time of day and category of each crime event. 
We analyse crime belonging to two broad categories of crime: violent and property crimes, which include homicides, sexual and non-sexual aggravated assaults, robbery, motor vehicle thefts and arson.
We assign each crime to a corehood through its position.

In order to estimate the number of crimes in a given core, we compute two types of features.
First, we consider the characteristics of the core itself. We include features that were previously found to attract potential offenders and targets~\cite{Wang:2016:CRI:2939672.2939736}, such as the \emph{residential population} and the number of \emph{nightlife}, \emph{shops} and \emph{food} POIs. 
Then, to account for the fact that environmental (neighbourhood) characteristic influence crime~\cite{jacobs1961death, sohn2016residential}, we consider corehood features in our model. We group them in Social disorganization (SD), Built Environment (BE) and Mobility (M) features.
The SD characteristics include the \emph{disadvantage}, \emph{instability} and \emph{ethnic diversity} of corehood. 
Consistently with the literature~\cite{Sampson, Sampson1989, Sampson1997, kubrin2003retaliatory}, \emph{disadvantage} and \emph{instability} are composite variables built from the two largest principal components of: (i) unemployment rate, (ii) poverty rate, defined as people living below the poverty line, and (iii) residential mobility rate, defined as the percentage of people who recently changed residency. 
Again, in accordance with the literature~\cite{sampson2013place, Sampson1989, sampson2009neighborhood}, \emph{ethnic diversity} is computed as the Hirschman-Herfindahl index across six population groups (e.g. hispanic, black, white people). Additional details are present in the Methods section. Note that we excluded all race-specific variables that are usually employed (e.g. percentage of black people) to build an evidence-based and race-neutral model.

The BE features are based on the Jane Jacobs theory~\cite{jacobs1961death}, which states that four conditions have to be valid to ensure a virtuous loop between the presence of people and a vibrant neighborhood life. First, a district should serve at least two or more functions to have streets continuously used by residents and strangers. Second, street blocks should be small and short to ensure both high \emph{walkability} and frequent meeting of people at street intersections. Third, diverse buildings make it possible to have low- and high-rent spaces, and thus a mixture of people and enterprises. The fourth condition is about dense concentration, which ensures a sufficient presence of people and enterprises to attract dwellers from different neighbourhoods continuously. This idea is summarized by the idea that "a well-used city street is apt to be a safe street and a deserted city street is apt to be unsafe"~\cite{jacobs1961death}. Moreover, \emph{walkability} is promotes social realtions~\cite{leyden2003social} and connected to local cohesion of neighbors.
Thus, in accordance with the literature~\cite{de2016death} we operationalize the four conditions in: i) \emph{land-use mix}; ii) \emph{block size} iii)  \emph{building age diversity}; iv)  \emph{population density} and \emph{walkability}, related to the second condition but also to density and reachability of POIs in the area.
The details of these metrics are available in the Methods section.

The M features are built upon recent mobility and criminology literature. We account for the average number of people at risk in the core by measuring the core \emph{ambient population}~\cite{andresen2006crime} and the \emph{attractiveness} of the corehood, where the latter is measured as the number of trips to the corehood for reasons different than travelling to work or home. 
\emph{Ambient population} and \emph{attractiveness} are computed by simulating realistic urban traces using Timegeo~\cite{Jiang13092016}, state of the art model for human mobility, in combination with mobile phone data. We do not include M features in Chicago, as we do not have mobile phone traces.

We model the relation of crime with core and corehood features through a spatially filtered Bayesian Negative Binomial, which is specifically tailored for discrete data, accounts for the overdispersion of crime events, and models uncertainty. 
The model accounts for the spatial auto-correlation, thus avoiding the biased parameters of non-spatial models~\cite{griffith2006spatial, tiefelsdorf2007semiparametric}. 
We identify spatial auto-correlation of crime events using a matrix indicating spatial proximity, and modelling spatial random effects.
Specifically, criminal activity is explained by a linear combination of an intercept, fixed effects (i.e. the input features), and random effects, which represent the unexplained variance that emerge from the spatial-autocorrelation of neighboring areas.
Although we find high spatial correlation in crime events, we did not find any significant spatial auto-correlation in the residuals with our spatial model (see Note 4 in the SI). 
The reader can refer to the Methods section for additional details about the model and its formulation.

\begin{figure*}[tbhp!]
	\centering
	\includegraphics[width=\textwidth]{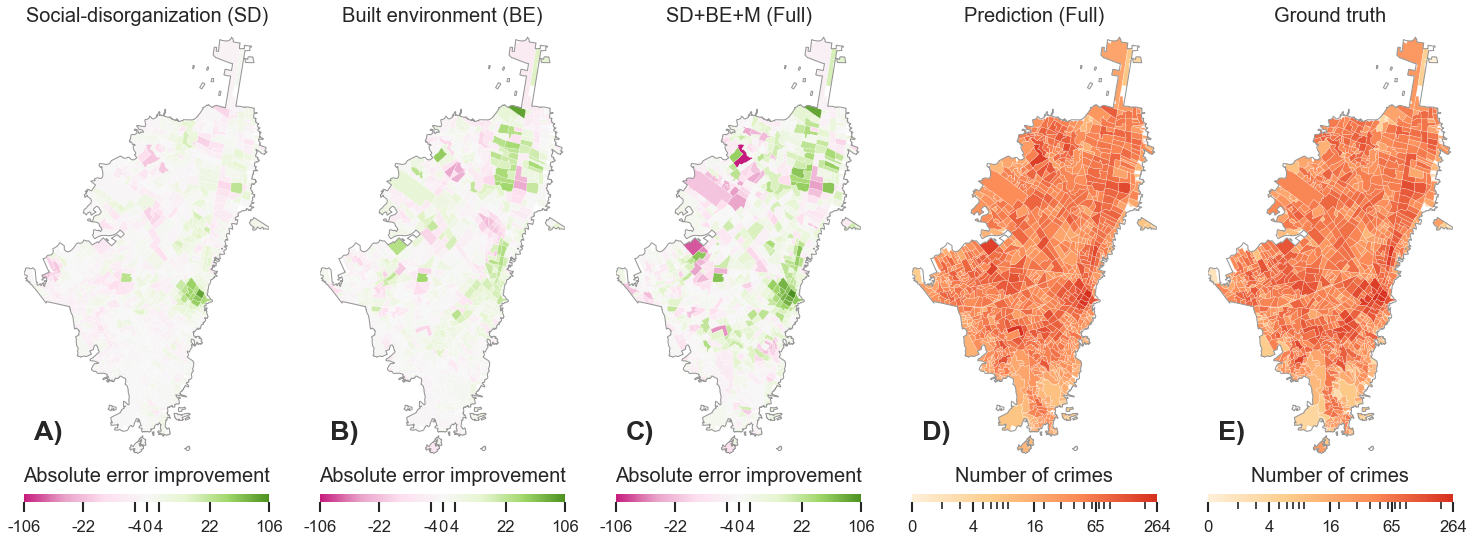}
	\caption{Maps of the estimated number of crime for each neighborhood in Bogotá for the A) Social-disorganization, B) Built environment, C) Full model. D) shows the Full model's prediction. E) shows the ground truth crime count.} 
	\label{fig:maps}
\end{figure*}

\begin{table*}[ht!]
	\centering
	\small
	\setlength{\tabcolsep}{4pt}
	\ra{1.2}
	\begin{tabular}{@{}lrr c rr c rr c rr@{}}
		\textbf{Model} &
		\multicolumn{2}{c}{\textbf{Bogotá}} && \multicolumn{2}{c}{\textbf{Boston}} && \multicolumn{2}{c}{\textbf{Los Angeles}} && \multicolumn{2}{c}{\textbf{Chicago}} \\
		\cmidrule{2-3} \cmidrule{5-6}  \cmidrule{8-9} \cmidrule{11-12} &
		$R^2_m$ ($R^2_c$) & LOO && $R^2_m$ ($R^2_c$) & LOO && $R^2_m$ ($R^2_c$) & LOO && $R^2_m$ ($R^2_c$) & LOO
		\\
		\midrule
		Core & 0.54 (0.75) & -3897 && 0.21 (0.64) & -2035 && 0.18 (0.68) & -9665 && 0.09 (0.68) & -8415 \\
		Social-disorganization (SD) & 0.57 (0.75) & -3891 && 0.55 (0.68) & -2019 && 0.53 (0.72) & -9529 && 0.66 (0.78) & -8019 \\
		Built environment (BE) & 0.61 (0.76) & -3881 && 0.36 (0.68) & -2014 && 0.27 (0.69) & -9629 && 0.21 (0.69) & -8371 \\
		Mobility (M) & 0.64 (0.80) & -3804 && 0.42 (0.70) & -2001 && 0.25 (0.70) & -9570 && - & - \\
		SD+BE & 0.64 (0.76) & -3881 && 0.65 (0.72) & -1987 && 0.56 (0.72) & -9508 && $\bst{0.67\ (0.79)}$ & $\bst{-8003}$ \\
		SD+M & 0.66 (0.81) & $\bst{-3795}$ && 0.67 (0.73) & -1973 && 0.55 (0.73) & -9467 && - & - \\
		BE+M & 0.68 (0.80) & -3819 && 0.50 (0.72) & -1989 && 0.30 (0.70) & -9585 && - & - \\
		SD+BE+M (Full) & $\bst{0.70\ (0.80)}$ & -3808 && $\bst{0.70\ (0.75)}$ & $\bst{-1957}$ && $\bst{0.56\ (0.74)}$ & $\bst{-9454}$ && - & - \\
		\bottomrule
	\end{tabular}
	\caption{Quantitative results of crime description and predictions in Bogotá, Boston, Los Angeles and Chicago. The model including Social Disorganization, Built Environment and Mobility features achieves the highest descriptive ($R^2_m$ and $R^2_c$) and predictive (LOO) performance. Here, we can see that contextual features of the neighborhood significantly increase our model's performance against the model considering only the core features.
		The LOO metric is calculated through the Pareto smoothed importance sampling Leave-One-Out cross-validation.}
	\label{table:DICresults}
\end{table*}

\subsection*{Description and prediction of crime}
For each city, we evaluate our model under various feature combinations to assess the contribution of each group of features. 
We measure the capability of the model to describe crime through the marginal $R^2_m$~\cite{nakagawa2017coefficient}, measuring the proportion of variance explained by the fixed effects (i.e. the input features). 
As reference, we also measure the conditional $R^2_c$~\cite{nakagawa2017coefficient} that takes into account both the variance explained by the fixed and random effects (i.e. the spatial autocorrelation) in explaining crime.
Additionally, we use the Pareto-smoothed importance sampling Leave-One-Out cross-validation (LOO)~\cite{vehtari2017practical} to assess the point-wise out-of-sample prediction accuracy (the higher, the better). 

First, we evaluate the baseline model that includes only the core variables. 
\Cref{table:DICresults} shows that the core-only model performs poorly in Chicago, Los Angeles and Boston, while it has high $R^2_m$ in Bogotá. The difference between $R^2_m$ and $R^2_c$ highlight that in all cities there is a significant unexplained variance that is captured by the spatial random effects, but not from the input features.

The SD, BE and M features significantly increase the explanatory power of our model. Particularly, in US cities, the $R^2_m$ increases up to 161\%, 194\% and 633\% in Boston, Los Angeles and Chicago.
Notably, and not surprisingly, the SD features are very important, especially in Chicago, were the "Chicago school" forged the Social Disorganization theory and further elaborated the role of collective efficacy on dealing with crime. 
Differently, the increase in Bogotá is less pronounced, suggesting that the neighbourhood impact on crime is limited.
Turning to M and BE features, we find that they describe the crime, but they are often as not meaningful as the SD features for crime prediction. 
However, the importance of mobility confirms the importance of floating population at describing microdynamic behaviour of criminal activity~\cite{caminha2017human}.
We observe that in all cities the conditional $R^2_c$ increases when adding the SD, BE and M features, revealing that the included variables also help explain the variance of crime across cores.

Overall, \Cref{table:DICresults} shows that considering together SD, BE and M variables result in the highest descriptive ($R^2_m$) and predictive (LOO) performance. 
This result means that, in order to model crime, one needs to account for multiple aspects of urban life, including Social Disorganization, the physical characteristics of the neighbourhoods, and mobility.
This result holds also against different combinations of the features (i.e. SD+BE, SD+M and BE+M). Nonetheless, some of the SD+BE and SD+M models are very competitive and might be considered when all data-sources are available.
Particularly, the ambient population (i.e. the average number of people who stop at the core) is one of the most important variables in the model and allows to better assess the number of people at risk, as suggested by previous works on aggregated mobility~\cite{caminha2017human}, satellite imagery~\cite{andresen2006crime}, Twitter~\cite{malleson2015spatio} and census data~\cite{Mburu2016}. 
However, we found that it might generate large errors due to places that are outliers of mobility in densely populated areas or hotspots of activity (see Figure S7 and Figure S8 in the SI).

$R^2_m$ improvements indicate that the model relies less on the random effects and it is better at explaining crime from the input features. 
\Cref{fig:maps} shows the spatial gain in performance from the baseline in Bogotá. 
First, it reveals that our Full model prediction resembles the ground truth data (\Cref{fig:maps} D-E), as confirmed by the high value of $R^2_c = 0.80$. 
Second, it shows that, while the SD and BE models achieve localized improvements (\Cref{fig:maps} A-B), the Full model improves the prediction almost everywhere. 
However, the Full model performs quite poorly in a specific area of Bogotá (see \Cref{fig:maps} C), part of the Engativá neighbourhood. 
By inspecting the coefficients of the model, we find that this area is an outlier as it is densely populated, thus resulting in an inflated prediction of crime, due to the high importance of residential and ambient population in the Bogotá model. Note, however, that our prediction is at the block level and the city-wide goodness of fit is $R^2_c = 0.80$.

The difference between $R^2_c$ and $R^2_m$ represents the unexplained variance due to spatial auto-correlation, which might suggest missing effects and variables. 
In Bogotá, our model points out that the touristic and dangerous neighbourhood La Candelaria, and the populous district of Engativá have significant unexplained variance that our input features cannot capture (see Figure S4 in the SI).
In Boston, the area near the Franklin park indicates missing local factors (see Figure S3 in SI). In Los Angeles, unexplained variance seems to be tied to places with a large number of people, namely the international airport and the UCLA campus (see Figure S5 in SI). Again, in Chicago, missing variables are suggested near the prison and the southern area (see Figure S6 in SI).
Altogether, these signals could help policymakers on including the best factors for each city and enacting policies that prevent crime.

Previous results suggested that the use of mobility flows between different regions might help describing crime~\cite{Wang:2016:CRI:2939672.2939736, wang2017region}. Thus, we test our model against this hypothesis by using the Origin-Destination matrix of people trips to model the auto-correlation between corehoods. The idea here is that human mobility might better explain the relation between corehoods than geographical closeness. However, we find that mobility flows significantly worsen the performance of our model (see Note 5 of SI).

While the effects of urban environment characteristics, socio-economic conditions, and mobility have been empirically tested separately~\cite{de2016death, graif2017neighborhood, lee_are_2017, sung2015residential, Sampson1997}, to the best of our knowledge, this is the first study to support with large-scale data the association of crime with socio-economic conditions, the built environment, and the mobility. 
However, we find that these aspects do not play the same role across cities, and only some of them contribute to the crime prediction model.

\begin{figure*}[hbt!]
	\centering
	\includegraphics[width=0.9\textwidth]{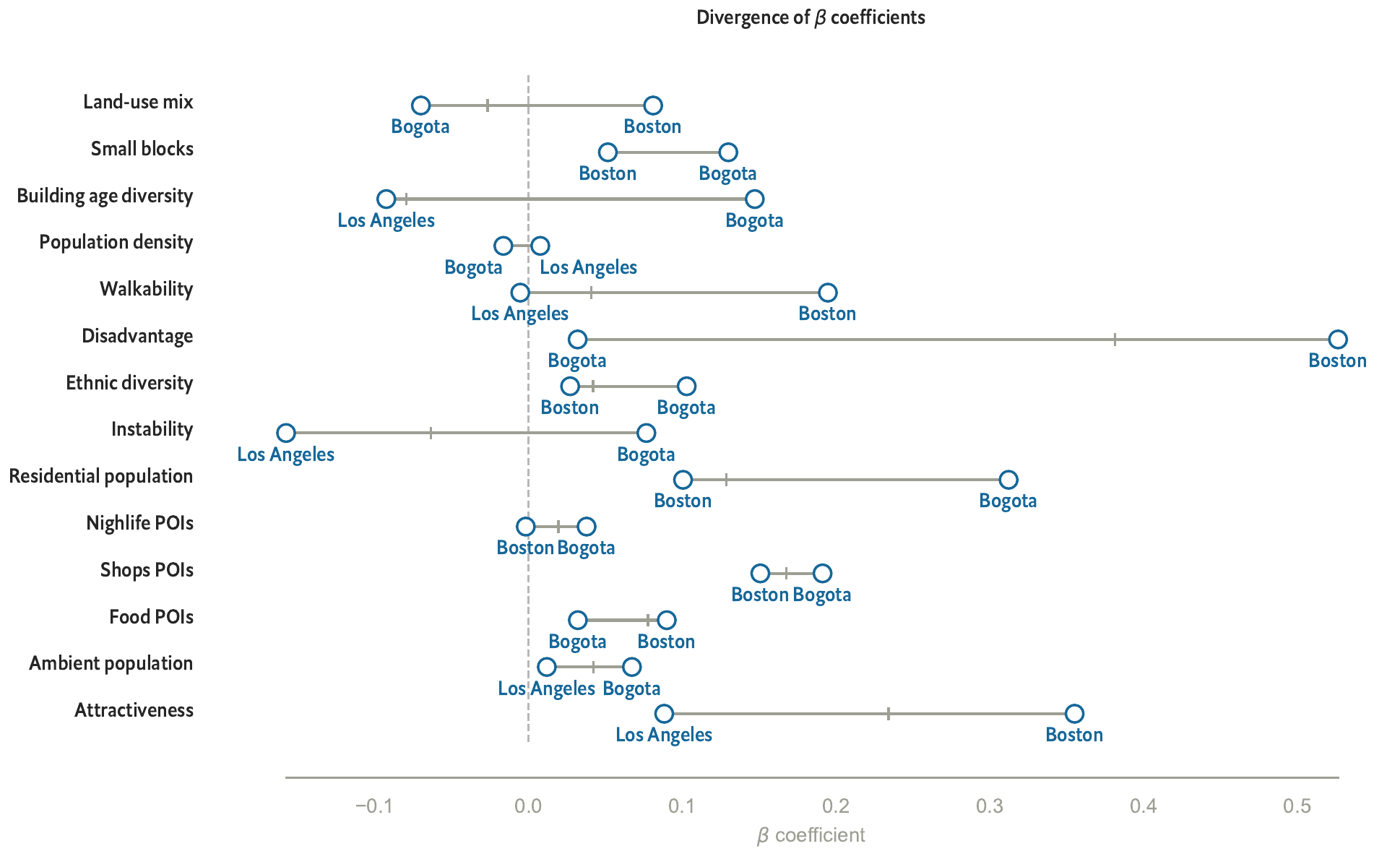}
	\caption{Generalized Linear Model's $\beta$ coefficients showing that Social Disorganization, Built Environment and Mobility features do not play the same role in all cities. We highlight in blue the minimum and maximum coefficient for each feature. Overall, this figure shows that there is no universal theory of crime.} 
	\label{fig:coefficientsPlot}
\end{figure*}

\subsection*{Neighborhood variables across cities}
In this section, we turn our attention to the standardized $\mathbf{\beta}$ coefficients that reveal how features correlate with criminal activity. 

First, we focus on the coefficients of the Full model, which combines socio-economic features with the characteristics of the built environment and human mobility. Note that here Chicago is excluded for lack of data.
\Cref{fig:coefficientsPlot} pictures that the $\mathbf{\beta}$ coefficients vary greatly across cities. 
For example, land-use mix correlates negatively with criminal activity in Bogotá and Los Angeles, but positively in Boston. Similarly, higher population building age diversity is present in low-crime areas in Boston and Los Angeles, but in high-crime areas in Bogotá.
Social disorganization variables are no less different, as corehood instability is correlated with crime activity only in Bogotá, differently from what expected from the theory~\cite{shaw1942juvenile, Sampson1989}.

The discrepancies between cities could be explained by the different spatial and socio-economic processes at play. 
When we look at the bivariate correlations across features, we observe interesting patterns. 
For example, in Los Angeles and Boston, \emph{walkability} is strongly positively correlated with population density and neighbourhood attractiveness, as expected~\cite{shaw1942juvenile, Sampson1989}, and slightly correlated with advantaged neighbourhoods. 
Differently, walkable areas in Bogotá have low population density areas and are highly advantaged, while the attractiveness is slightly correlated (see Figure S11 in SI). 
A possible reason for the $\mathbf{\beta}$ coefficients disagreement lies on the multi-collinearity of the input features. Although we use the QR decomposition and Ridge penalty to shrink down the variables that are not necessary, the difference between the coefficients is present also in simpler models. 

The difference between the results across cities also suggests that crime correlates differently with space and people. For example, we observe that in Bogotá high crime areas relate to advantaged neighbourhoods, while in Boston and Los Angeles higher crime seem to be linked to disadvantaged neighbourhoods, according to the theory~\cite{shaw1942juvenile, Sampson1989}. A possible explanation might be related to under-reporting and police disrespecting, which seems to be a problem particularly in Bogotá~\cite{godoy2018security}. However, literature has shown how neighbourhood cultural codes, informal local control, and problematic policing are also related to violent criminal activities~\cite{kubrin2003retaliatory}.

However, some features behave similarly in all the cities. We find that corehoods with high disadvantage and ethnic diversity but, surprisingly, smaller blocks have higher crime activity. 
While in the core we find that the presence of Shops, Food POIs, and population (both residential and ambient) correlates positively with criminal activity. These results resonate with literature showing that the presence of POIs and ambient population increase crime due to a higher number of potential targets and offenders in an area. Additionally, we find that corehood attractiveness has a strong connection with crimes, suggesting that the presence of people that do not live nor work in the area might influence crime. This result is in contrast with literature based on Jacobs' theory~\cite{jacobs1961death, traunmueller2014mining}, but resonate with Oscar Newman's one arguing that a high number of visitors results in higher anonymity and, thus, crime~\cite{newman1972defensible}. Additionally, a recent empirical study from survey data~\cite{boivin2018crimes} agrees with our result, obtained instead with large-scale and passively collected information.
In the supplementary materials (SI), we compare all the cities in detail.

To test the possibility of having a universal model that predicts crime, we test a model that uses only the features that behave in the same direction in all the cities. This model consistently performs worse than the Full model (see Note 10 in SI), showing that at this moment, no model is convenient to be easily applied to all cities. 
We also studied at what extent a model trained in one city can be tested to another city. We found that US cities are, as expected, more similar to each other than Bogotá, and that Los Angeles behave similarly to Chicago.

\section*{Discussion}
 In this paper, we modelled the presence of crime across four cities, widely different with respect to cultural, economic, historical and geographical aspects. 
 We found that the variability of the dynamics and history of each city poses a challenge to the existence of a model that "fits it all", able to learn from one city and to predict on another one. 
Instead, we presented a model that could describe and disentangle the role of diverse factors in urban crime and draw some theoretical and practical implications.  

The goal of this research goes beyond crime prediction in time (i.e. forecasting). Offences are concentrated in a small number of places~\cite{Lee2017}, and are tightly coupled with places, stable over time~\cite{weisburd2012criminology}. Thus, the easiest way to predict crime is modelling those few places with the highest number of crimes, also known as \emph{hotspots}~\cite{bogomolov2014once, short2010dissipation}. On the contrary, we seek to shed light on the diverse set of factors at play with urban crime and do predictions for those areas without crime statistics (i.e. nowcasting). 

Our cumulative results show little evidence in support of the Jane Jacobs' theory, arguing that specific urban features and people on the street generate higher security. On the contrary, we often found that Jacobs' features and urban vibrancy increase people's vulnerability to crime, suggesting that further work has to be done in this direction.

We found that different theories often seen as competing can complement each other in models that take into account the socio-economic, built environment and mobility conditions together. 
The importance of mobility and built environment characteristics showed that competitive descriptive and predictive models can be built from data available at large scale without the necessity of costly in-field survey studies. 
However, we found that aspects related to social disorganisation are important for crime description and prediction. Therefore, it is crucial to consider alternative sources of data to infer social cohesion and interactions and overcome the use of census information, which is costly to collect and rarely updated. There have been multiple attempts at inferring social interactions~\cite{eagle2009inferring}, poverty~\cite{Blumenstock1073}, well-being~\cite{Pappalardo2016} and unemployment~\cite{Toole2015} but so far very little work has been done at micro spatial levels.

Comparing multiple cities in different countries do not come without limitations. First, our analysis ignore temporal variation such as opening times of POIs or temporal variation in mobility. 
Second, due to lack of consistent data, we did not account for variables such as political and housing policies, security perception, community participation, and social ties within family and within neighbourhoods that were previously found to be related to crime~\cite{faust2019social, salesses2013collaborative, tran2013participation}. 
Finally, official crime data do not come without errors, given that not all crimes are reported nor recorded~\cite{small2018understanding}, and there is no "ground truth" data to gauge any bias in police records. 

Our work seeks to make headway on the previous limitation of a single site of study origin. While recent works have started the use of street units and blocks to study criminal activity~\cite{contreras2017block, hipp2019effect, kim2020street, rosser2017predictive}, they often relied on a small subset of variables and one city. 
Analysing multiple cities together exposed criminology theories to discrepancies and differences. 
Descriptive modelling can help policymakers to understand the use of urban space and deploy future investments and resources thoughtfully.
Moreover, from the scientific perspective, descriptive modelling can provide insights for strong predictors, and potentially for explanatory variables, to be further investigated by explanatory modelling and experiments~\cite{doi:10.1146/annurev-statistics-031017-100204}.
Thus, we hope that additional research keeps exploring multi-dimensional aspects related to crime, to clarify potential crime causes and design better cities.

\section*{Methods}
The socio-economical and Jane Jacobs' urban theories are dependent upon the actions and activities at work in communities. Thus, we identified corehoods as social and geographical units of analysis.
Then, we obtained and aggregated the data for each corehood of Bogotá, Boston, Los Angeles and Chicago.
	
\subsection*{Crime data} Data collection mechanisms and crime categories can vary from country to country. The Uniform Crime Reporting (UCR) Program (\url{https://ucr.fbi.gov/}) is a US statistical effort to make crime reports uniform across the country. The UCR divides crime in two main groups: Part 1 and Part 2 offences. The former is composed by violent crimes (aggravated assault, forcible rape, robbery and murder) and property crimes (larceny-theft, motor vehicle theft, burglary and arson), while the latter are considered less serious and they include offences such as simple assaults and nuisance crimes. For each city we thus collect the geo-referenced data of committed crimes and we filter out those crimes not belonging to Part 1 of UCR, similarly to most of the criminology literature. We categorized crimes in Bogotá consistently with UCR categories and released the mapping for future comparisons. 
We reference crimes to cores and, when a crime event happens in a street segment shared between cores, we evenly assign the event to both cores. Due to the limit in accuracy of GPS positioning, we create a buffer of 30 meters for each crime, which is the distance usually employed for stop location detection algorithms~\cite{de2019strategies}.
More details are presented in the SI. We summed crime events over one year to minimize seasonal fluctuations.
    
\subsection*{Mobile phone data}
We computed the ambient population and the OD matrices for Bogotá, Boston and Los Angeles through the TimeGeo modelling framework~\cite{Jiang13092016}. We fitted the model starting from aggregated and anonymized Call Detailed Records (CDRs) collected from 12-01-2013 to 05-31-2014, 6 weeks in 2010, and  10-15, 2012 to 11-24, 2012 for Bogotá, Boston and Los Angeles respectively. The anonymized data for the three cities was collected for billing purposes by two mobile operators, who also kindly provided to us the data for the present research.
Timegeo is an agent-based model that simulates the activity of people from mobile phone data. To be consistent with the travel surveys of each city it simulates the time, duration, direction and type of travels within the city. The types of travels are classified as Home-Based from/to Work (HBW), Home-Based from/to Other type of locations (HBO) and Non-Home-based from/to Other type of locations (NHB). To build the \emph{ambient population} we counted the number of people who stops at a specific location for at least one hour, while we built the corehood \emph{attractiveness} counting the number of NHB trips with the corehood as destination.

\subsection*{Spatial and census data}
Census blocks, population, employment and poverty for US cities were drawn from the American Community Survey (ACS) (\url{https://www.census.gov/programs-surveys/acs}). For US cities we also used some city-specific datasets that are described in the SI.
The census data of Bogotá was obtained by the Departmento Administrativo Nacional de Estadística (DANE), which organized the 2005 general census for the city (\url{http://www.dane.gov.co}). The poverty data of Bogotá was extracted from the Sisbén in the Identification System III of 2014.
The detailed list of datasets and URLs are listed in the SI.

\subsection*{Built environment features} We operationalize the Jane Jacobs conditions through some state of the art metrics defined in literature~\cite{de2016death}.
The land-use mix is computed as the average entropy among land uses: $\text{LUM}_{L,i} = - \sum_{j \in L} \frac{P_{i,j} \log(P_{i,j})}{\log(|L|)}$, where $P_{i,j}$ is the percentage of square meters having land use $j$ in unit $i$, and $L = \{\text{residential}, \text{commercial and institutional},$ $\text{park and recreational}\}$ represents the considered land uses in the metric. 
The LUM ranges between 0, wherein the unit is composed by only one land use (e.g. residential), and 1, wherein developed area is equally shared among the $n$ land-uses.
    
Then, for each corehood we determine the \emph{walkability} through the accessibility of the core to the nearest point of interests (e.g. convenience stores, restaurants, sport facilities). Consistently with literature~\cite{walkscore}, we define the weighted \emph{walkability} score as: $\text{walk}_i = \frac{1}{|B_i|} \sum_{c \in C} \sum_{b \in B_i} \wdist(b, \closest(b, \text{POI}_c))$, where $C$ is the set of categories (i.e., Food, Shops, Grocery, Schools, Entertainment, Parks and outside, Coffee, Banks, Books), $\wdist$ is the street-network distance decay function, and $\text{POI}_c$ is the set of POIs of category $c$.
The distance decay function gives a weight (importance) to each POI reachable from a staring point. Additional information about the \emph{walkability} score can be find in the SI.

We then compute the average block area among the set $B_i$ of blocks in unit $i$ as $\text{Blocks area}_i = \frac{1}{|B_i|} \sum_{b \in B_i} \area(b)$, and the building age diversity as the standard deviation of building ages in the corehood. 
    
Finally, we operationalize Jacobs' density condition with the dwelling units density, computed from census data. Additional details are described in the SI.

\subsection*{Social-disorganization}
We create the feature \emph{disadvantage} and \emph{instability} through the two largest PCA principal components of: (i) unemployment rate, (ii) poverty rate, defined as people living below the poverty line, and (iii) residential mobility rate, defined as the percentage of people who recently changed residency (one year for US cities and fiver years for Bogotá). From the loadings of the PCA linear combination we verified that disadvantage is mainly a linear combination of poverty rate and unemployment, while instability is mainly about residential mobility rate.

In the Social-disorganization variables we do not include any ethnic-specific variables (e.g. percentage of black people) other than diversity because they might be present only in some places and not in others (e.g. native Americans in Bogotá), and to avoid any ethnic-specific bias. Ethnic diversity represents the difficulties of a community to communicate and collaborate for a common goal. Accordingly to the literature, it is computed as the Hirschman-Herfindahl diversity index of six population groups $H = 1- \sum_{i=1}^N s_i^2$, where $s_i$ is the proportion of people belonging to the ethnicity $i$, and $N$ is the number of ethnicities. Consistently with the literature we include for US cities: Hispanics, non-Hispanic Blacks, Whites, Asians, Native Hawaiians - Pacific Islanders and others. For Bogotá we include: Indigenous, Rom, Islanders (San Andrés), Palenquero, Black and others. 

\subsection*{Bayesian model}
    Let $y_i$ be the discrete number of crimes for a set of spatial regions $i=  ,\ldots, N$. 
    We approximate the relation between crimes and spatial features through a Negative Binomial approach that models the non-negative nature of the crime-counts in a city, but also the overdispersion found in the data (Note 4 in the SI). Specifically, $\log(\mathbb{E}(Y)) = \mathbf{X\beta} + \mathbf{b}$  where $\mathbf{X}$ is the input data and $\mathbf{\beta}$ the coefficients of the model.
    $\mathbf{b}$ are the random effects that accounts for the unexplained variability of crime (i.e. the spatial-autocorrelation). In this paper, we account the spatial auto-correlation with the Bayesian Spatial Filtering (BSF)~\cite{hughes2017spatial} that defines $\mathbf{b} = \mathbf{E}\mathbf{\gamma}$ where $\mathbf{\gamma}$ are coefficients to be found. 
    $\mathbf{E}$ is instead defined as the first principal components of $\mathbf{E} = \mathbf{MCM}$, where $\mathbf{C}$ is a spatial matrix that describes the graph between spatial locations, while $\mathbf{M} = \mathbf{I} - \mathbf{X}(\mathbf{X}'\mathbf{X})-\mathbf{X}'$, which is an approximation of the spatial error model~\cite{tiefelsdorf2007semiparametric}. We tested for the presence of spatial auto-correlation on the residuals of all the models without finding significant auto-correlation.
	As the results might change with different definitions of $\mathbf{C}$, we tested all the models for three definitions: i) $\mathbf{C}$ is a binary adjacency matrix identifying whether a corehood overlaps another corehood, ii) $\mathbf{C}$ is a inverse distance matrix between corehoods, iii) $\mathbf{C}$ describes the flow of people between corehoods, which is extracted from mobile phone data. We found that the binary matrix consistently outperforms other definitions.
    Additional details of the presented models, definition of $\mathbf{C}$, and other competitive models tested are present in the SI. 
    
    As we have to account for collinearity, we employ a Ridge penalty to all fixed effects.

    \subsection*{Model calibration ed evaluation} Model calibration is carried out by means of Markov Chain Monte Carlo (MCMC) approach. Convergence was assured by the Gelman-Rubin convergence statistics, and discarding the first 15,000 iterations and running the model over 5,000 iterations. 
    
     We assess how well the models describe crime through the conditional $R^2$ and the marginal $R^2$~\cite{nakagawa2017coefficient}, which adapt the popular coefficient of determination to the generalized linear mixed-effects models. They are defined as:
    \begin{align*}
    R^2_m &= \frac{\sigma_f^2}{\sigma_f^2 + \sigma_r^2 + \sigma_{\epsilon}^2}\\
    R^2_c &= \frac{\sigma_f^2 + \sigma_r^2}{\sigma_f^2 + \sigma_r^2 + \sigma_{\epsilon}^2}
    \end{align*}
    where $\sigma_f^2$ is the variance explained by the fixed effects, $\sigma_r^2$ is the variance explained by the random effects, and $\sigma_{\epsilon}^2$ is the variance of the residuals. 
    Specifically, $f= \mathbf{X}\beta$, $r= \mathbf{E}\gamma$ and $\epsilon$ is specific to the Negative Binomial and defined~\cite{nakagawa2017coefficient} as $\epsilon= \ln{(1+1/\mu+1/\phi)}$, with $\mu = \frac{1}{N} \sum_i^N y_i$ and $\phi$ is the shape parameter of the Negative Binomial distribution.
    
    We assess the out of sample predictive accuracy through the Pareto-smoothed importance sampling Leave-One-Out cross-validation (LOO)~\cite{vehtari2017practical}, which is the state of the art for evaluating Bayesian models.

\section*{Data Availability}
We are pleased to make available the source-code and datasets accompanying this research. The projects files are available at \url{https://github.com/denadai2/bayesian-crime-multiple-cities/}.

\section*{Acknowledgements}
We thank Paolo Bosetti and Junpeng Lao for the helpful comments. We especially thank Andrés Clavijo for his support on the data, we all hope that this work could make Bogotá better.
This work was supported by the Berkeley DeepDrive and the ITS Berkeley 2018-19 SB1 Research Grant (to M.C.G.); the French Development Agency and the World Bank (to M.D.N., B.L. and E.L.).

\section*{Author contributions statement}
M.D.N, E.L., M.C.G. and B.L. designed research and experiments; M.D.N, Y.X., M.C.G. and B.L. performed research and experiments; M.D.N, M.C.G. and B.L. contributed new analytic tools; M.D.N, and Y.X. analysed the data; and M.D.N, M.C.G. and B.L. wrote the paper. All authors read, reviewed and approved the final manuscript.

\section*{Competing Interests}

The authors declare no competing interests.

\bibliography{bibliography}
\clearpage

\appendix
\section{Walkability} \label{label1}
\label{walkability}
We determine the \emph{walkability} of a neighbourhood through its accessibility to the nearest Point Of Interests (\emph{e.g.}, convenience stores, restaurants, sport facilities). 
The concept of \emph{walkability} is empirically calculated in many different ways. 
However, one of the most accepted one is Walk Score~\cite{walkscore}. We here describe and compute the \emph{walkability} score for our cities consistently with their methodology~\cite{walkscore}, as Walk Score is not available for all the cities we consider.

Thus, for each city block $b$ we first collect an ordered list of $n_c$ closest Point Of Interests (POIs) belonging to category $c$:
\begin{equation}
    \text{closest}(b, c) = \left[p_1, p_2, \ldots, p_{n_c} \right]
\end{equation}
where $p_1$ is the closest POI of category $c$ to $b$, $p_2$ is the second closest and so on so forth.
And then we compute the \emph{walkability} score as:
\begin{equation}
\text{walk}_i = \sum_{c \in C} \sum_i^{n_c} w_{c, i} \distance(b, \text{closest}(b, c)_i)
\label{eq:walkability}
\end{equation}
where $C$ is the set of categories (i.e. Food, Shops, Grocery, Schools, Entertainment, Parks and outside, Coffee, Banks, Books), $\distance$ is the street-network distance decay function (explained later), and $w_{c,i}$ is a weighting factor that depends on both the category $c$ and the $i$-est closest POI.

In categories where depth of choice is important, multiple POIs are considered (i.e. $n_c > 1$). 
For example, restaurants and bars are combined in a single category due to their overlapping function. 
They are the most frequent walking destination, hence we include 10 counts of places to account for the depth of offer in the neighbourhood. The shopping category represent all the retails where people can buy products such as clothes, gifts, etc. 
They are common walking destinations and they are commonly described as important for the attractiveness of a place. Thus, we considered 5 counts of places for this category. Coffee shops are also important for the neighbourhood, but not as important as restaurants and shopping places. Thus, we considered 2 counts for this category. For other categories only the distance from the nearest POI is calculated. 
These parameters are consistent with Walk Score~\cite{walkscore}.
The definitions of $w$ and $n_c$ as summarized in \Cref{table:walkscore1}.

The amenities are extracted from Foursquare, a crowd-sourced project where people participate in an online game where they check-in places where they go.

\begin{table*}[tbhp!]
	\centering
	\footnotesize
		\begin{tabular}{@{}lrl@{}}
			\textbf{Category} & $\mathbf{n_c}$ & $\mathbf{w}$ \\
			\midrule
			Grocery & 1 & $\left[ 3 \right]$\\
			Food & 10 & $\left[ .75,.45,.25,.25,.225,.225,.225,.225,.2,.2 \right]$\\
			Shops & 5 & $\left[.5,.45,.4,.35,.3\right]$\\
            Schools & 1 & $\left[1\right]$\\
            Entertainment & 1 & $\left[1\right]$\\
            Parks and outside & 1 & $\left[1\right]$\\
            Coffee & 2 & $\left[1.25,.75\right]$\\
            Banks & 1 & $\left[1\right]$\\
            Books & 1 & $\left[1\right]$\\
			\bottomrule
		\end{tabular}
		\caption{Additional details to compute the \emph{walkability} score. $n_c$ is the number of retrieved Point of Interests, while $w$ is an ordered sequence of weights applied to the distances to the nearest $n_c$ Point of Interests.}
	\label{table:walkscore1}
\end{table*}

The distance decay function that computes importance weight to each POI reachable from a starting point. Similarly to Walk Score, we use a polynomial distance that assigns the maximum score to amenities $\sim 500$ meters far from the starting point, then the score decays quickly until 1500 meters, where it first slows down then it goes to zero. The distance is along the street network, instead of the geometric distance (see \Cref{fig:eff}).

\begin{figure}[t!]
	\centering
	\includegraphics[width=0.5\columnwidth]{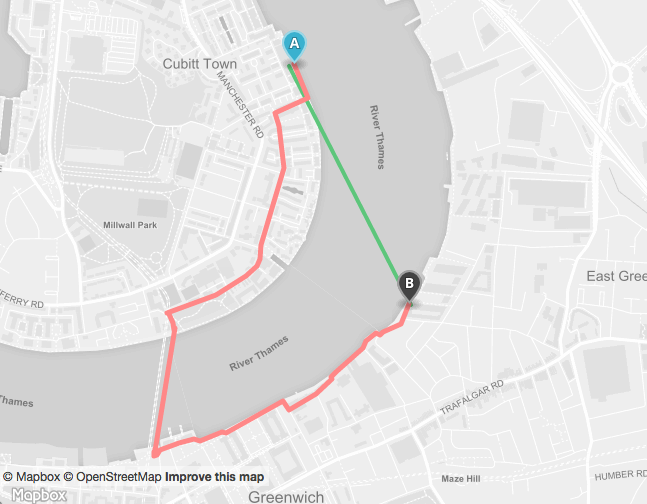}
	\caption{The geometric distance (green) and shortest path walking distance (red) between two points.} 
	\label{fig:eff}
\end{figure}

\section{Crime measure}

In criminology, there are usually two ways to assess crime: crime counts and rates. The main difference between the two consists on how the population at risk is modelled. The former assumes that the importance of the population at risk has to be found by the model, while the latter assumes that importance is equal to one. Specifically, in crime rates model, the crime counts with the population at risk, which is usually the residential one.
Recently, some scholars have discussed the costs and benefits of alternative denominators such as the ambient population~\cite{andresen2006crime}. Different ways exist on how to compute it, including the use of satellite imagery~\cite{andresen2006crime}, census data~\cite{Mburu2016} and Foursquare check-ins~\cite{kadar2017measuring}.

However, it is not clear whether using residential rates, ambient rates, nor the policy implications of using one or the other. Moreover, the bias of ambient rates can be potentially lead to misleading interpretation of crime. In this research, we thus prefer to describe crime counts controlling for residential and ambient population. Hence, it is also possible to describe their relative role through the $\beta$ coefficients of the regression.

\section{Data sources}

\begin{table*}[tbhp!]
	\centering
	\footnotesize
		\begin{tabular}{@{}llcccl@{}}
			\textbf{Type} & \textbf{City} & \textbf{Open Data} & \multicolumn{2}{c}{\textbf{Data is shared}} \\
			\cmidrule{4-5} &&& Raw & Aggregated \\
			\midrule
			\multirow{4}{*}{Blocks} & 
			    Bogotá &  & \checkmark & \checkmark \\
			    &Boston & \checkmark & \checkmark & \checkmark \\
			    &Chicago & \checkmark & \checkmark & \checkmark \\
			    &LA & \checkmark & \checkmark & \checkmark \\
			\midrule
			\multirow{4}{*}{Census Blocks} & 
			    Bogotá & \checkmark\cite{data_bogota_ideca} & \checkmark & \checkmark \\
			    &Boston & \checkmark\cite{data_UStiger} & \checkmark & \checkmark \\
			    &Chicago & \checkmark\cite{data_UStiger} & \checkmark & \checkmark \\
			    &LA & \checkmark\cite{data_UStiger} & \checkmark & \checkmark \\
			\midrule
			\multirow{4}{*}{Boundaries} & 
			    Bogotá &  & \checkmark & \checkmark \\
			    &Boston & \checkmark\cite{data_BostonMaps} & \checkmark & \checkmark \\
			    &Chicago & \checkmark\cite{data_chicagomaps} & \checkmark & \checkmark \\
			    &LA & \checkmark\cite{data_LAmaps} & \checkmark & \checkmark \\
			\midrule
			\multirow{4}{*}{Buildings} & 
			    Bogotá & \checkmark\cite{data_bogota_buildings} & \checkmark & \checkmark \\
			    &Boston & \checkmark\cite{data_boston_buildings} & \checkmark & \checkmark \\
			    &Chicago & \checkmark\cite{data_chicago_buildings} & \checkmark & \checkmark \\
			    &LA & \checkmark\cite{data_LA_buildings} & \checkmark & \checkmark \\
			\midrule
			\multirow{4}{*}{Crime} & 
			    Bogotá &  & & \checkmark \\
			    &Boston & \checkmark\cite{data_boston_crime} & \checkmark & \checkmark \\
			    &Chicago & \checkmark\cite{data_chicago_crime} & \checkmark & \checkmark \\
			    &LA & \checkmark\cite{data_LA_crime} & \checkmark & \checkmark \\
			\midrule
			\multirow{4}{*}{Employment and Ethnic mix} & 
			    Bogotá &  & & \checkmark \\
			    &Boston & \checkmark\cite{data_USfact} & \checkmark & \checkmark \\
			    &Chicago & \checkmark\cite{data_USfact} & \checkmark & \checkmark \\
			    &LA & \checkmark\cite{data_USfact} & \checkmark & \checkmark \\
			\midrule
			\multirow{4}{*}{Land Use} & 
			    Bogotá & \checkmark\cite{data_bogota_ideca} & \checkmark & \checkmark \\
			    &Boston & \checkmark\cite{data_boston_landuse} & \checkmark & \checkmark \\
			    &Chicago & \checkmark\cite{data_chicago_landuse} & \checkmark & \checkmark \\
			    &LA & \checkmark\cite{data_LA_landuse} & \checkmark & \checkmark \\
			\midrule
			Mobile phone data & All but Chicago &  & & \checkmark \\
			\midrule
			POIs & All &  & & \checkmark \\
			\midrule
			\multirow{4}{*}{Population} & 
			    Bogotá &  & & \checkmark \\
			    &Boston & \checkmark\cite{data_USfact} & \checkmark & \checkmark \\
			    &Chicago & \checkmark\cite{data_USfact} & \checkmark & \checkmark \\
			    &LA & \checkmark\cite{data_USfact} & \checkmark & \checkmark \\
			\midrule
			\multirow{4}{*}{Poverty} & 
			    Bogotá &  & & \checkmark \\
			    &Boston & \checkmark\cite{data_USfact} & \checkmark & \checkmark \\
			    &Chicago & \checkmark\cite{data_USfact} & \checkmark & \checkmark \\
			    &LA & \checkmark\cite{data_USfact} & \checkmark & \checkmark \\
			\midrule
			\multirow{4}{*}{Residential stability} & 
			    Bogotá &  & & \checkmark \\
			    &Boston & \checkmark\cite{data_USfact} & \checkmark & \checkmark \\
			    &Chicago & \checkmark\cite{data_USfact} & \checkmark & \checkmark \\
			    &LA & \checkmark\cite{data_USfact} & \checkmark & \checkmark \\
			\midrule
			Street network & All & \checkmark & \checkmark & \checkmark 
			
			\\
			\bottomrule
		\end{tabular}
		\caption{Description of the data sources to replicate the paper. Most of the data is shared along this paper in raw format, while a subset of it (POIs and mobile phone data) could be shared only in an aggregated format for license and privacy reasons.}
	\label{supplementary:table-datasets}
\end{table*}

\subsection*{Distribution of crime}
\begin{figure}[t!]
	\centering
	\includegraphics[width=0.7\columnwidth]{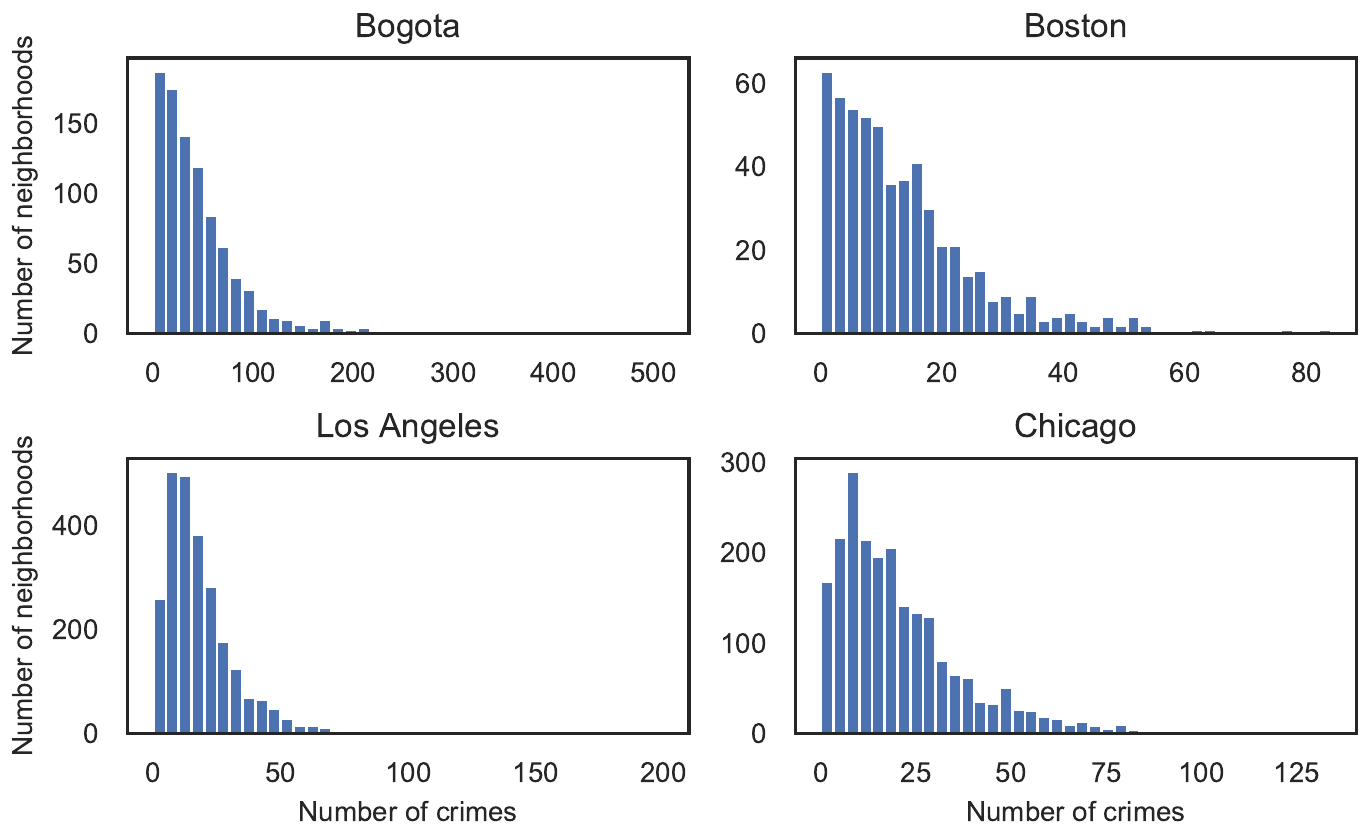}
	\caption{Distribution of the number of total committed crimes for each neighborhood.} 
	\label{fig:distributionCrime}
\end{figure}

Crime is not evenly distributed in space and its distribution in the neighbourhoods of the analysed cities can be seen in~\Cref{fig:distributionCrime}. 

\section{The spatial model}

In each city, we tested the presence of spatial auto-correlation of crime $y$ through the Moran's I coefficient~\cite{moran1948interpretation} $MC(y)$. When $MC(y) > 0$ there is positive auto-correlation, negative otherwise. 
In the former case, places with high crime tend to be near places with high crime, in the latter places with high crime are near places with low crime. 
When $MC$ is not near zero, regression models might exhibit spatial correlation in the residuals, thus invalidating the assumption of independence of the errors. In these cases, regression models should account for the spatial auto-correlation between spatial units, as we did.

In our paper, we model the crime counts in a city with a Negative Binomial (NB) regression, and we account for the spatial auto-correlation with the Bayesian Spatial Filtering (BSF)~\cite{hughes2017spatial} approach.
The NB model is defined as:
\begin{equation}
    \text{NegBinomial2}(y \, | \, \mu, \phi)  = \binom{y + \phi - 1}{y} \,
\left( \frac{\mu}{\mu+\phi} \right)^{\!y} \, \left(
\frac{\phi}{\mu+\phi} \right)^{\!\phi} \!
\label{eq:spatial_model}
\end{equation}
where the mean and variance of $y$ are: $\mathbb{E}[Y] = \mu \text{ and }\text{Var}[Y] = \mu + \frac{\mu^2}{\phi}$.
We use the logarithm as link function for the NB.

The BSF is defined as:
\begin{eqnarray*}
    \log(\mathbb{E}[Y]) &=& \beta_0 + \bm{X}\beta + \mathbf{E}\mathbf{\gamma}\\
	\mathbf{\gamma} | \rho &\sim& \mathcal{N}(\mathbf{0}, (\rho \mathbf{E^\intercal QE})^{-1})\\
	\rho &\sim& \Gamma(0.5, 2000).
\end{eqnarray*}
where $\mathbf{Q}$ is the Laplacian of $\mathbf{W}$, and $\rho$ is a Gamma with a large mean to discourage artifactual spatial structures in the posterior~\cite{arnold1999bayesian, hughes2017spatial}.
Since we want to account for the correlation between the features, and well generalize the model, we apply a Ridge penalty to the $\beta$ coefficients and the QR decomposition to decorrelate covariates and, thus, the resulting posterior distribution.
Thus, we model $\beta_0$ and $\beta$ as:
\begin{eqnarray*}
    \beta_0 &\sim& \mathcal{N}(0,1)\\
    \beta | \tau &\sim& \mathcal{N}(0, \tau)\\
	\tau &\sim& \mathcal{C}^{+}(0, 1).
\end{eqnarray*}

The alternative formulation that does not account for the auto-correlation is a NB model with Ridge penalty on the beta coefficients, which is defined as:
\begin{equation*}
    \log(\mathbb{E}[Y]) = \beta_0 + \bm{X}\beta
\end{equation*}
where $C^{+}(0, 1)$ is the half-Cauchy distribution with a mean of zero and a scale parameter of one. We chose the half-Cauchy as suggested by Andrew Gelman~\cite{gelman2006prior}.

In \Cref{table:residual_MC} we show the NB model exhibits a strong positive spatial auto-correlation in the residuals, while the BSF model does not, as expected. The models based on BSF are also superior in the LOO and $R^2_{c}$.

\subsection*{Test for overdispersion} The NB model is motivated by the extra-Poisson variability of the crime distribution in the city. We can test the need of overdispersion through the Potthoff-Whittinghill and the Lagrange multiplier test.

The Potthoff-Whittinghill index of dispersion test~\cite{potthoff2006homogeneity} rejects the hypothesis of no overdispersion. It is defined as:
\begin{equation}
\sum\limits_{i=1}^n (y_i-\bar{y})^2 / \bar{y}
\end{equation}
which is approximately a chi-square distribution with $k-1$ degrees of freedom. 
We also apply the Lagrange multiplier test, defined as:
\begin{equation}
\frac{(\sum_{i=1}^n \mu_i^2 - n\bar{y})^2}{2 \sum_{i=1}^n \mu_i^2}
\end{equation}
With one degree of freedom, the test appears to be significant – the hypothesis of no overdispersion is again rejected.

\subsection*{Selection of E eigenvectors}
Following seminal literature of eigen-based spatial modelling and filtering~\cite{tiefelsdorf2007semiparametric}, we select the first $k$ eigenvectors from $\mathbf{E} = \mathbf{MCM}$, where $\mathbf{C}$ is a spatial matrix that describes the graph between spatial locations, while $\mathbf{M} = \mathbf{I} - \mathbf{X}(\mathbf{X}'\mathbf{X})-\mathbf{X}'$, which is an approximation of the spatial error model.

The associated sets of eigenvalues ($\lambda_1, \lambda_2, \ldots, \lambda_n$) from the $\mathbf{MCM}$ decomposition assess the strength of a spatial pattern. 
A vector $i$ with $\lambda_i > 0$ describes positive spatial auto-correlation, while vectors with $\lambda_i < 0$ describe negative spatial auto-correlation. 
Spatial models are notoriously inefficient at dealing with negative spatial auto-correlation, which is also rather rare. Thus, consistently with literature~\cite{chun2016eigenvector} we focus on positive auto-correlation and we select those vectors having $\lambda' = \lambda/\lambda_{max} >= 0.25$, where $\lambda_{max}$ is the maximum value among the eigenvalues.

\subsection*{Test for residual auto-correlation}
We test for the presence of auto-correlation in the models' residuals to the Moran's I. However, since our model is not an Ordinary Least Squares, we used a corrected version of Moran's I that is specifically tailored for log-linear relationships~\cite{lin2007loglinear}. The index is defined as:
\begin{equation*}
	I_{p} = \frac{n}{\sum_{i,j = 0}^n \mathbf{W}_{i,j}} \frac{r' \mathbf{W} r}{r'r}
\end{equation*}
where $n$ is the number of spatial units, $\mathbf{W}$ is the spatial matrix, and $r$ is the vector of residual errors.

We do not find significant residual spatial auto-correlation in our spatial models.

\begin{table*}[tbhp!]
	\centering
    \footnotesize
	\ra{1.2}
	\begin{tabular}{@{}llrrrrrrrrrrrrrr@{}}
		\textbf{City} & \textbf{Model}  &
		\multicolumn{3}{c}{\textbf{BSF}} & \phantom{a} & \multicolumn{3}{c}{\textbf{NB}} \\
		\cmidrule{3-5} \cmidrule{7-9}  &&
		LOO & $R^2_c$ & $MC_p$ && LOO & $R^2_c$ & $MC_p$
		\\
		\midrule
		
		\multirow{1}{*}{Bogota} & Core & \textbf{-3897} & \textbf{0.75} & \textbf{-0.034} && -4126 & 0.53 & 0.455 \\
		& Social-disorganization (SD) & \textbf{-3891} & \textbf{0.75} & \textbf{-0.043} && -4079 & 0.58 & 0.354 \\
		& Built environment (BE) & \textbf{-3881} & \textbf{0.76} & \textbf{-0.036} && -4061 & 0.61 & 0.371 \\
		& Mobility (M) & \textbf{-3804} & \textbf{0.80} & \textbf{-0.042} && -4034 & 0.64 & 0.460 \\
		& SD+BE & \textbf{-3880} & \textbf{0.76} & \textbf{-0.035} && -4013 & 0.65 & 0.287 \\
		& SD+M & \textbf{-3795} & \textbf{0.81} & \textbf{-0.050} && -3988 & 0.67 & 0.374 \\
		& BE+M & \textbf{-3819} & \textbf{0.80} & \textbf{-0.025} && -3980 & 0.68 & 0.361 \\
		& SD+BE+M (Full) & \textbf{-3809} & \textbf{0.80} & \textbf{-0.040} && -3941 & 0.71 & 0.284 \\
		\midrule
		\multirow{1}{*}{Boston} & Core & \textbf{-2035} & \textbf{0.64} & \textbf{-0.005} && -2209 & 0.22 & 0.418 \\
		& Social-disorganization (SD) & \textbf{-2019} & \textbf{0.68} & \textbf{-0.003} && -2088 & 0.55 & 0.236 \\
		& Built environment (BE) & \textbf{-2014} & \textbf{0.68} & \textbf{-0.033} && -2169 & 0.37 & 0.309 \\
		& Mobility (M) & \textbf{-2001} & \textbf{0.70} & \textbf{-0.026} && -2140 & 0.45 & 0.351 \\
		& SD+BE & \textbf{-1987} & \textbf{0.72} & \textbf{-0.043} && -2030 & 0.65 & 0.108 \\
		& SD+M & \textbf{-1973} & \textbf{0.73} & \textbf{-0.030} && -2011 & 0.67 & 0.105 \\
		& BE+M & \textbf{-1989} & \textbf{0.72} & \textbf{-0.033} && -2109 & 0.52 & 0.264 \\
		& SD+BE+M (Full) & \textbf{-1957} & \textbf{0.75} & \textbf{-0.040} && -1993 & 0.70 & 0.084 \\
		\midrule
		\multirow{1}{*}{LA} & Core & \textbf{-9665} & \textbf{0.68} & \textbf{0.032} && -10757 & 0.17 & 0.647 \\
		& Social-disorganization (SD) & \textbf{-9529} & \textbf{0.72} & \textbf{0.005} && -10042 & 0.55 & 0.416 \\
		& Built environment (BE) & \textbf{-9629} & \textbf{0.69} & \textbf{0.005} && -10618 & 0.27 & 0.615 \\
		& Mobility (M) & \textbf{-9570} & \textbf{0.70} & \textbf{0.018} && -10658 & 0.24 & 0.628 \\
		& SD+BE & \textbf{-9508} & \textbf{0.72} & \textbf{-0.010} && -9989 & 0.57 & 0.366 \\
		& SD+M & \textbf{-9467} & \textbf{0.73} & \textbf{-0.002} && -10003 & 0.57 & 0.444 \\
		& BE+M & \textbf{-9585} & \textbf{0.70} & \textbf{0.011} && -10571 & 0.30 & 0.618 \\
		& SD+BE+M (Full) & \textbf{-9453} & \textbf{0.74} & \textbf{-0.011} && -9967 & 0.58 & 0.388 \\
		\midrule
		\multirow{1}{*}{Chicago} & Core & \textbf{-8415} & \textbf{0.68} & \textbf{0.117} && -9350 & 0.09 & 0.543 \\
		& Social-disorganization (SD) & \textbf{-8019} & \textbf{0.78} & \textbf{0.016} && -8391 & 0.66 & 0.295 \\
		& Built environment (BE) & \textbf{-8371} & \textbf{0.69} & \textbf{0.093} && -9237 & 0.21 & 0.519 \\
		& SD+BE & \textbf{-8003} & \textbf{0.79} & \textbf{0.003} && -8357 & 0.68 & 0.282 \\
\bottomrule
	\end{tabular}
	\caption{Comparison between the BSF model and the NB model that does not account for the spatial auto-correlation.}
	\label{table:residual_MC}
\end{table*}

\section{Alternative spatial models}
We tested alternative spatial models that could explain the residual spatial-autocorrelation. Here, we compare the BSF with other two similar, but competitive models: the Random Effects Eigenvector Spatial Filtering (RE-ESF)~\cite{murakami2019eigenvector} and the Linear ESF model~\cite{tiefelsdorf2007semiparametric}.
The ESF model is defined as:
\begin{eqnarray*}
    log(\mathbb{E}[Y]) &=& \alpha + \bm{X}\beta + \mathbf{E}\gamma\\
	\gamma | \rho &\sim& \mathcal{N}(0, \rho \lambda)\\
	\rho^{-2} | \nu &\sim& \Gamma(\nu/2, \nu/2)\\
	\nu &\sim& \Gamma(2, 0.1).
\end{eqnarray*}
where $\lambda$ is the vector $L \times 1$ of the eigenvalues associated with $\mathbf{E}$, and $\rho$ is chosen to constrain the spatial random effects $\gamma$ and avoid they penalize too much the fixed effects. To ensure limited variance, $v$ is limited to an upper value of 2.

The RE-ESF instead assumes $\mathbf{\gamma}$ to be random such that:
\begin{eqnarray*}
	\mathbf{\gamma} | \alpha, \rho &\sim& \mathcal{N}(0, \rho \mathbf{\Lambda}(\omega))\\
	\omega^{-1} &\sim& \Gamma(2, 5).
\end{eqnarray*}
where $\lambda(\omega) = \frac{\sum_l \lambda_l}{\sum_l \lambda_l^\omega} \lambda_l^\omega$ is a multiplier that represents the scale of spatial variance, and $\omega$ is a parameter to be found.

\Cref{table:alternativeModels} shows that no models clearly outperforms another, suggesting that they are almost equivalent in a Full Bayesian setting.

\begin{table*}[tbhp!]
	\centering
    \footnotesize
	\ra{1.2}
	\begin{tabular}{@{}llrrrrrrrrrrrrrr@{}}
		\textbf{City} & \textbf{Model}  &
		\multicolumn{2}{c}{\textbf{BSF}} & \phantom{a} & \multicolumn{2}{c}{\textbf{RE-ESF}} & \phantom{a} & \multicolumn{2}{c}{\textbf{ESF}} \\
		\cmidrule{3-4} \cmidrule{6-7} \cmidrule{9-10} &&
		LOO & $MC_p$ && LOO & $MC_p$ && LOO & $MC_p$
		\\
		\midrule
		
		\multirow{1}{*}{Bogota} & Core & \textbf{-3897} & -0.034 && -3899 & -0.045 && -3902 & -0.041\\
		& Social-disorganization (SD) & \textbf{-3891} & -0.043 && -3895 & -0.052 && -3896 & -0.049\\
		& Built environment (BE) & \textbf{-3881} & -0.036 && -3882 & -0.045 && -3884 & -0.042\\
		& Mobility (M) & -3804 & -0.042 && \textbf{-3803} & -0.048 && -3807 & -0.046\\
		& SD+BE & \textbf{-3880} & -0.035 && -3882 & -0.043 && -3884 & -0.040\\
		& SD+M & \textbf{-3795} & -0.050 && -3796 & -0.057 && -3798 & -0.056\\
		& BE+M & -3819 & -0.025 && \textbf{-3817} & -0.033 && -3822 & -0.032\\
		& SD+BE+M (Full) & \textbf{-3809} & -0.040 && -3810 & -0.049 && -3810 & -0.046\\
		\midrule
		\multirow{1}{*}{Boston} & Core & \textbf{-2035} & -0.005 && \textbf{-2035} & -0.016 && \textbf{-2035} & -0.014\\
		& Social-disorganization (SD) & \textbf{-2019} & -0.003 && -2020 & -0.017 && -2020 & -0.016\\
		& Built environment (BE) & -2014 & -0.033 && \textbf{-2013} & -0.044 && -2014 & -0.044\\
		& Mobility (M) & -2001 & -0.026 && \textbf{-1999} & -0.035 && \textbf{-1999} & -0.035\\
		& SD+BE & -1987 & -0.043 && -1987 & -0.057 && \textbf{-1986} & -0.057\\
		& SD+M & -1973 & -0.030 && -1972 & -0.046 && \textbf{-1971} & -0.045\\
		& BE+M & -1989 & -0.033 && \textbf{-1988} & -0.043 && \textbf{-1988} & -0.042\\
		& SD+BE+M (Full) & -1957 & -0.040 && -1957 & -0.054 && \textbf{-1956} & -0.053\\
		\midrule
		\multirow{1}{*}{LA} & Core & -9665 & 0.032 && \textbf{-9663} & 0.028 && -9671 & 0.029\\
		& Social-disorganization (SD) & \textbf{-9529} & 0.005 && -9530 & -0.001 && -9535 & -0.000\\
		& Built environment (BE) & \textbf{-9629} & 0.005 && \textbf{-9629} & 0.002 && -9638 & 0.001\\
		& Mobility (M) & -9570 & 0.018 && \textbf{-9569} & 0.014 && -9576 & 0.014\\
		& SD+BE & \textbf{-9508} & -0.010 && -9510 & -0.015 && -9514 & -0.014\\
		& SD+M & \textbf{-9467} & -0.002 && -9468 & -0.006 && -9472 & -0.006\\
		& BE+M & \textbf{-9585} & 0.011 && \textbf{-9585} & 0.008 && -9591 & 0.007\\
		& SD+BE+M (Full) & \textbf{-9453} & -0.011 && -9455 & -0.015 && -9458 & -0.015\\
		\midrule
		\multirow{1}{*}{Chicago} & Core & -8415 & 0.117 && \textbf{-8413} & 0.115 && -8414 & 0.115\\
		& Social-disorganization (SD) & \textbf{-8019} & 0.016 && \textbf{-8019} & 0.012 && \textbf{-8019} & 0.013\\
		& Built environment (BE) & -8371 & 0.093 && \textbf{-8369} & 0.090 && -8370 & 0.090\\
		& SD+BE & \textbf{-8002} & 0.003 && -8004 & -0.001 && -8006 & -0.000\\
\bottomrule
	\end{tabular}
	\caption{Results of alternative spatial models applied in each city.}
	\label{table:alternativeModels}
\end{table*}

\section{Alternative Connectivity Matrices}
Incorporating spatial relationship in spatial models requires the definition of a connectivity matrix $\mathbf{C}$ that describes the relationship (if any) between one spatial unit and all the others.
One of the most common connectivity matrix is a binary relationship between spatial units, also called topology representation~\cite{griffith2006spatial}, which usually results in a sparse matrix:
\begin{equation}
  \mathbf{C} = (c_{i,j}) = \left\{
  \begin{array}{@{}ll@{}}
    1, & \text{if}\ i\text{ is neighbour of } j \text{ and } i \neq j \\
    0, & \text{otherwise}
  \end{array}\right.
\end{equation} 
An alternative formulation is based on distance. For example, Griffith \emph{et al.}~\cite{griffith2006spatial} defines:
\begin{equation}
  \mathbf{C} = (c_{i,j}) = \left\{
  \begin{array}{@{}ll@{}}
    0, & \text{if}\ i = j \\
    0, & \text{if}\ d > t \\
    1-(d_{i,j}/4t)^2, & \text{if}\ d \leq t
  \end{array}\right.
\end{equation}
where $t$ is chosen as the maximal distance that keeps all the spatial units connected, while $d_{i,j}$ is the (Euclidian) distance between the centroids of unit $i$ and $j$. $t$ is computed through the maximal distance of a Minimum Spanning Tree (MST) computed on the distance matrix $\mathbf{D}$.

We also test for an additional formulation that accounts for the connectivity of spatial units, extracted from mobile phone data. Here, the assumption is that the more connections two sites have, the strongest is the similarity between them. Thus $c_{i,j} = t_{i,j}$, where $t$ is the number of trips made, on average, from the unit $i$ to unit $j$ and viceversa. As the matrix $\mathbf{T}$ is not symmetrical and it does not have the diagonal equal to zero:
\begin{equation}
    \mathbf{T} = (\mathbf{T} + \mathbf{T}^\intercal)/2
\end{equation}
and $t_{i,j}=0$ for all $i=j$.

As shown in \Cref{table:alternative_connectivity}, the BSF model with contiguity matrix achieves better performance in all the urban settings.

\begin{table*}[tbhp!]
	\centering
    \footnotesize
	\ra{1.2}
	\begin{tabular}{@{}llrrrrrrrrrrrrrr@{}}
		\textbf{City} & \textbf{Model}  &
		\multicolumn{2}{c}{\textbf{Contiguity}} & \phantom{a} & \multicolumn{2}{c}{\textbf{Distance}} & \phantom{a} & \multicolumn{2}{c}{\textbf{Mobility}} \\
		\cmidrule{3-4} \cmidrule{6-7} \cmidrule{9-10} &&
		LOO & $MC_p$ && LOO & $MC_p$ && LOO & $MC_p$
		\\
		\midrule
		
		\multirow{1}{*}{Bogota} & Core & \textbf{-3897} & -0.034 && -3999 & 0.016 && -4104 & 0.073\\
		& Social-disorganization (SD) & \textbf{-3891} & -0.043 && -3969 & 0.010 && -4051 & 0.060\\
		& Built environment (BE) & \textbf{-3881} & -0.036 && -3971 & 0.003 && -4051 & 0.048\\
		& Mobility (M) & \textbf{-3804} & -0.042 && -3906 & 0.010 && -4021 & 0.037\\
		& SD+BE & \textbf{-3880} & -0.035 && -3949 & 0.006 && -4000 & 0.042\\
		& SD+M & \textbf{-3795} & -0.050 && -3879 & 0.002 && -3964 & 0.031\\
		& BE+M & \textbf{-3819} & -0.025 && -3889 & 0.003 && -3975 & 0.027\\
		& SD+BE+M (Full) & \textbf{-3809} & -0.040 && -3873 & 0.001 && -3929 & 0.027\\
		\midrule
		\multirow{1}{*}{Boston} & Core & \textbf{-2035} & -0.005 && -2078 & 0.036 && -2208 & 0.118\\
		& Social-disorganization (SD) & \textbf{-2019} & -0.003 && -2037 & 0.024 && -2081 & 0.107\\
		& Built environment (BE) & \textbf{-2014} & -0.033 && -2040 & 0.010 && -2168 & 0.076\\
		& Mobility (M) & \textbf{-2001} & -0.026 && -2014 & -0.010 && -2094 & 0.029\\
		& SD+BE & \textbf{-1987} & -0.043 && -2007 & -0.002 && -2029 & 0.022\\
		& SD+M & \textbf{-1973} & -0.030 && -1991 & -0.011 && -2011 & 0.001\\
		& BE+M & \textbf{-1989} & -0.033 && -2006 & -0.006 && -2108 & 0.075\\
		& SD+BE+M (Full) & \textbf{-1957} & -0.040 && -1976 & -0.008 && -1993 & -0.002\\
		\midrule
		\multirow{1}{*}{LA} & Core & \textbf{-9665} & 0.032 && -9966 & 0.061 && -10691 & 0.134\\
		& Social-disorganization (SD) & \textbf{-9529} & 0.005 && -9708 & 0.014 && -9977 & 0.033\\
		& Built environment (BE) & \textbf{-9629} & 0.005 && -9767 & 0.017 && -10528 & 0.113\\
		& Mobility (M) & \textbf{-9570} & 0.018 && -9875 & 0.050 && -10596 & 0.118\\
		& SD+BE & \textbf{-9508} & -0.010 && -9668 & 0.005 && -9922 & 0.026\\
		& SD+M & \textbf{-9467} & -0.002 && -9647 & 0.016 && -9898 & 0.029\\
		& BE+M & \textbf{-9585} & 0.011 && -9733 & 0.021 && -10514 & 0.111\\
		& SD+BE+M (Full) & \textbf{-9453} & -0.011 && -9613 & 0.003 && -9891 & 0.027\\
		\midrule
		\multirow{1}{*}{Chicago} & Core & \textbf{-8415} & 0.117 && -8716 & 0.076 && - & -\\
		& Social-disorganization (SD) & \textbf{-8019} & 0.016 && -8257 & 0.028 && - & -\\
		& Built environment (BE) & \textbf{-8371} & 0.093 && -8623 & 0.058 && - & -\\
		& SD+BE & \textbf{-8003} & 0.003 && -8244 & 0.032 && - & -\\
\bottomrule
	\end{tabular}
	\caption{Results of alternative connectivity matrices applied in the spatial model of each city. Chicago does not have mobility information, thus it was not possible to use the mobility matrix.}
	\label{table:alternative_connectivity}
\end{table*}

\section{Spatial model decomposition}
The fit of our model can be decomposed in fixed effects, which are the input variables, random effects, which are the unexplained variance through spatial auto-correlation, and residuals, which are the errors of the model. 

\begin{figure*}[h!]
	\centering
	\includegraphics[width=0.9\textwidth]{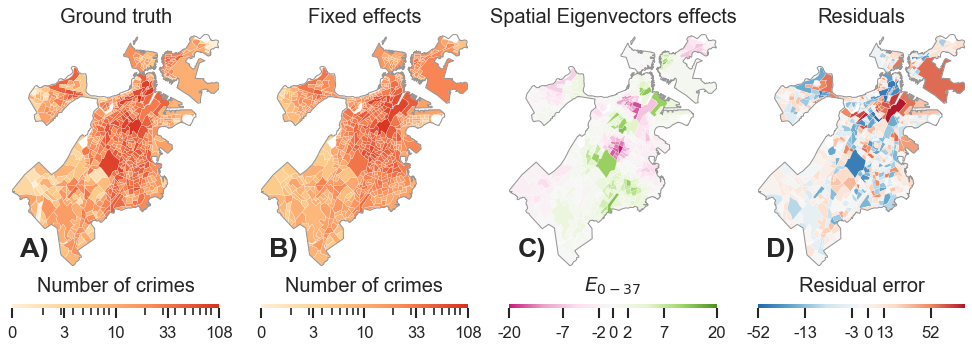}
	\caption{Decomposition of the ground truth in fixed, random and residuals effects in Boston.} 
	\label{fig:boston_map}
\end{figure*}

\begin{figure*}[h!]
	\centering
	\includegraphics[width=0.9\textwidth]{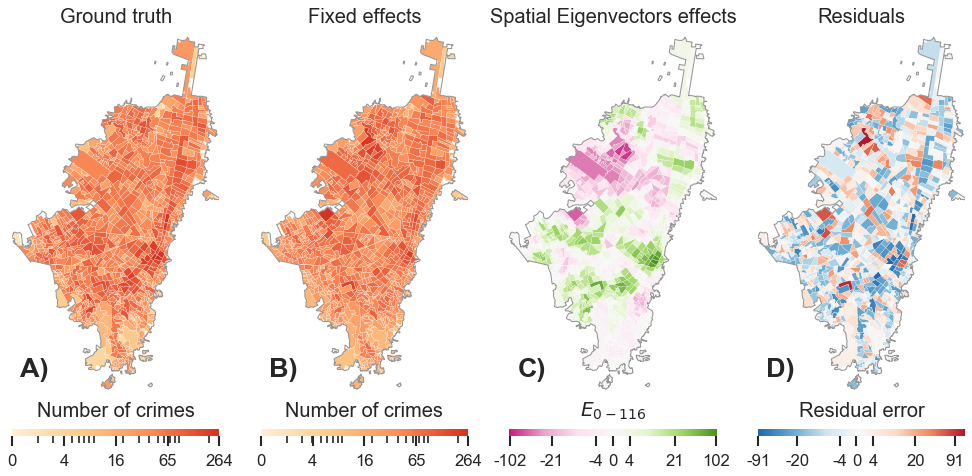}
	\caption{Decomposition of the ground truth in fixed, random and residuals effects in Bogotá.} 
	\label{fig:bogota_map}
\end{figure*}

\begin{figure*}[h!]
	\centering
	\includegraphics[width=0.9\textwidth]{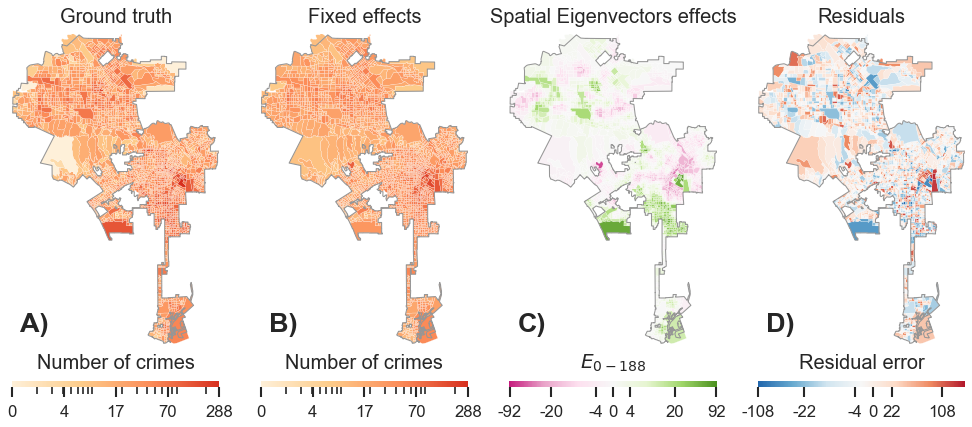}
	\caption{Decomposition of the ground truth in fixed, random and residuals effects in Los Angeles.} 
	\label{fig:LA_map}
\end{figure*}

\begin{figure*}[h!]
	\centering
	\includegraphics[width=0.9\textwidth]{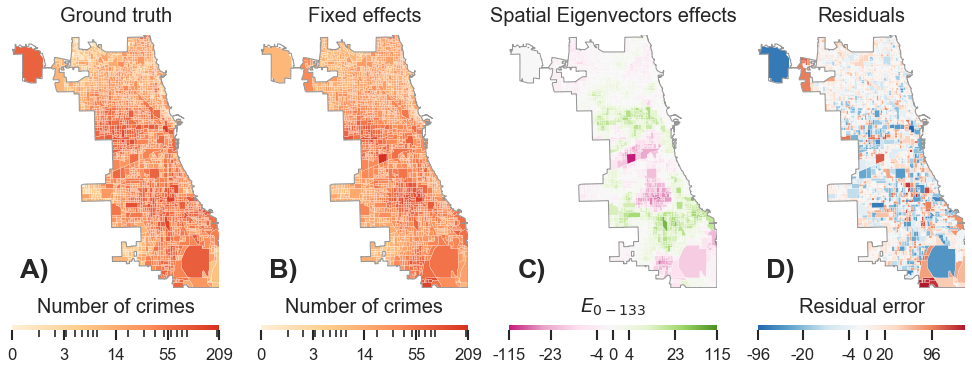}
	\caption{Decomposition of the ground truth in fixed, random and residuals effects in Chicago.} 
	\label{fig:chicago_map}
\end{figure*}

From \Cref{fig:boston_map} D, \Cref{fig:bogota_map} D, \Cref{fig:LA_map} D, and \Cref{fig:chicago_map} D we do not observe any clear spatial pattern on the residuals, confirming that the BSF model is easing the spatial auto-correlation as expected.

The observation of the random effects can help on locating local spatial effects that are not considered in the fixed effects. In Bogotá, the model suggests that significant unexplained variance is present near the touristic and dangerous neighbourhood  La Candelaria, and near the populous district of  Engativá (see \Cref{fig:bogota_map}). In Boston, the area near the Franklin park indicates missing local factors (see \Cref{fig:boston_map}). In Los Angeles, unexplained variance seems to be tied to places with a large amount of people, namely the international airport and the UCLA campus (see SI \Cref{fig:LA_map}). Finally in Chicago missing variables are suggested near the prison and the southern area (see \Cref{fig:chicago_map}).

\section{Improvement analysis}
\begin{figure*}[h!]
	\centering
	\includegraphics[width=0.9\textwidth]{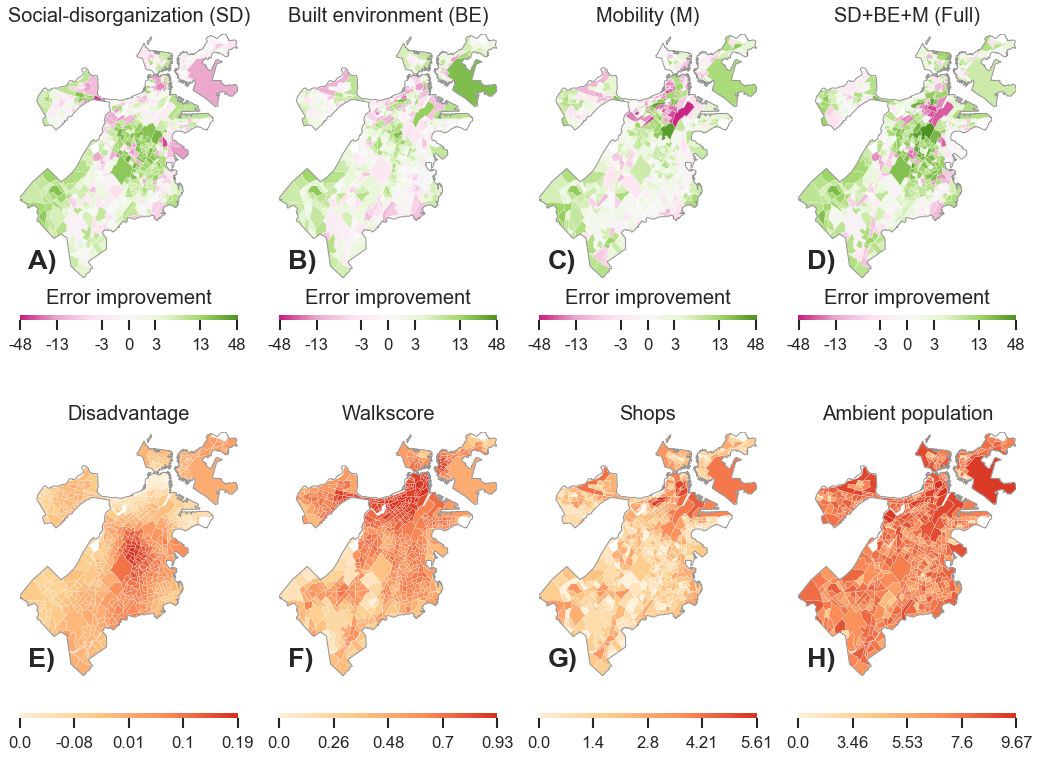}
	\caption{Model improvements in Boston. It can be seen that SD model improves the prediction from the core model almost everywhere but especially in disadvantaged areas, while the BE model seems to better improve the prediction near the airport and peripheral areas. The mobility model seems to improve but it also generates a strong outlier that performs poorly near the city centre. Finally, the Full model outperforms the core model and the other models almost everywhere. It keeps failing in some areas due to mobility information.} 
	\label{fig:boston_improvement}
\end{figure*}

\begin{figure*}[h!]
	\centering
	\includegraphics[width=0.9\textwidth]{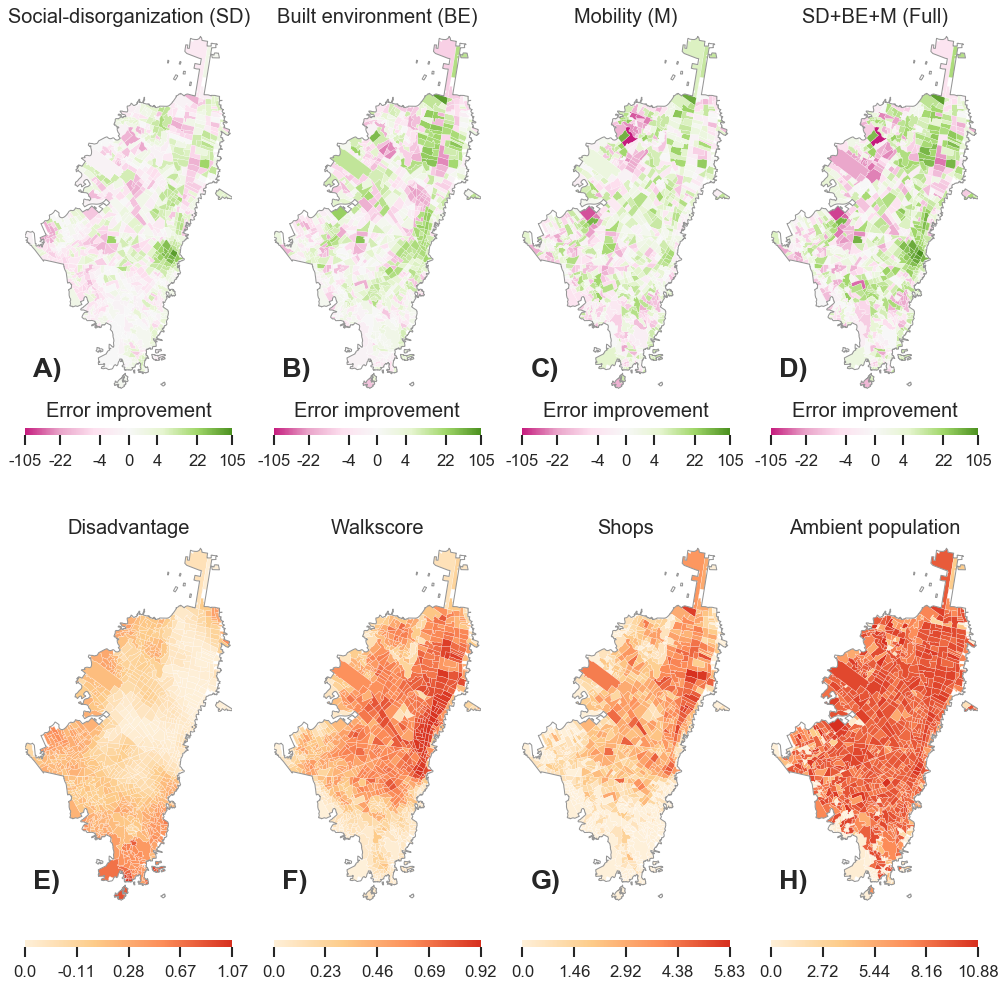}
	\caption{Model improvements in Bogotá. It can be seen that SD model improves the prediction from the core model very slightly, while the BE model seems to better improve the prediction near the richer part of the city and in areas with high number of shops. The mobility model seems to improve but it also generates a strong outlier that performs poorly near the  ``Engativa", a populous neighbourhood in Bogotá. Finally, the Full model outperforms the core model and the other models almost everywhere. It keeps failing in some areas due to mobility information.}
	\label{fig:bogota_improvement}
\end{figure*}

\begin{figure*}[h!]
	\centering
	\includegraphics[width=0.9\textwidth]{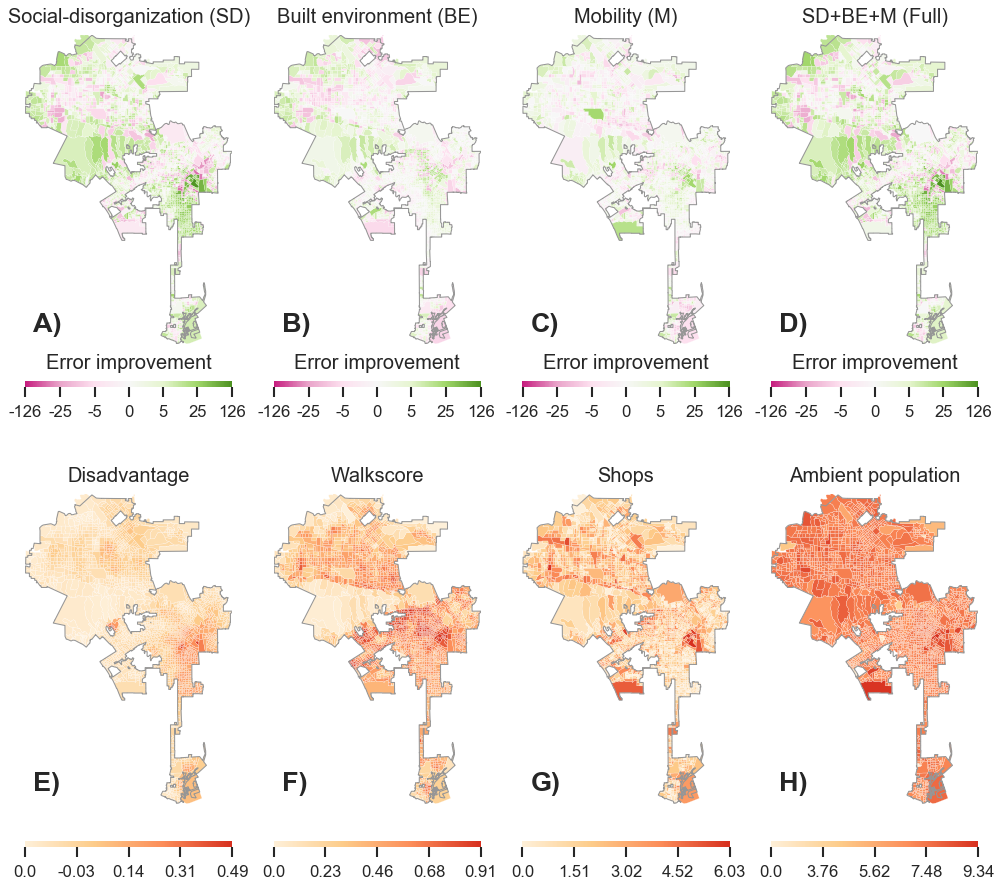}
	\caption{Model improvements in Los Angeles. It can be seen that SD model improves the prediction from the core model very consistently, especially in deprived areas, while the BE model seems to only slightly improve the prediction. The mobility model seems to improve especially in popular areas, such the airport. Finally, the Full model outperforms the core model and the other models almost everywhere.}
	\label{fig:LA_improvement}
\end{figure*}

\begin{figure*}[h!]
	\centering
	\includegraphics[width=0.9\textwidth]{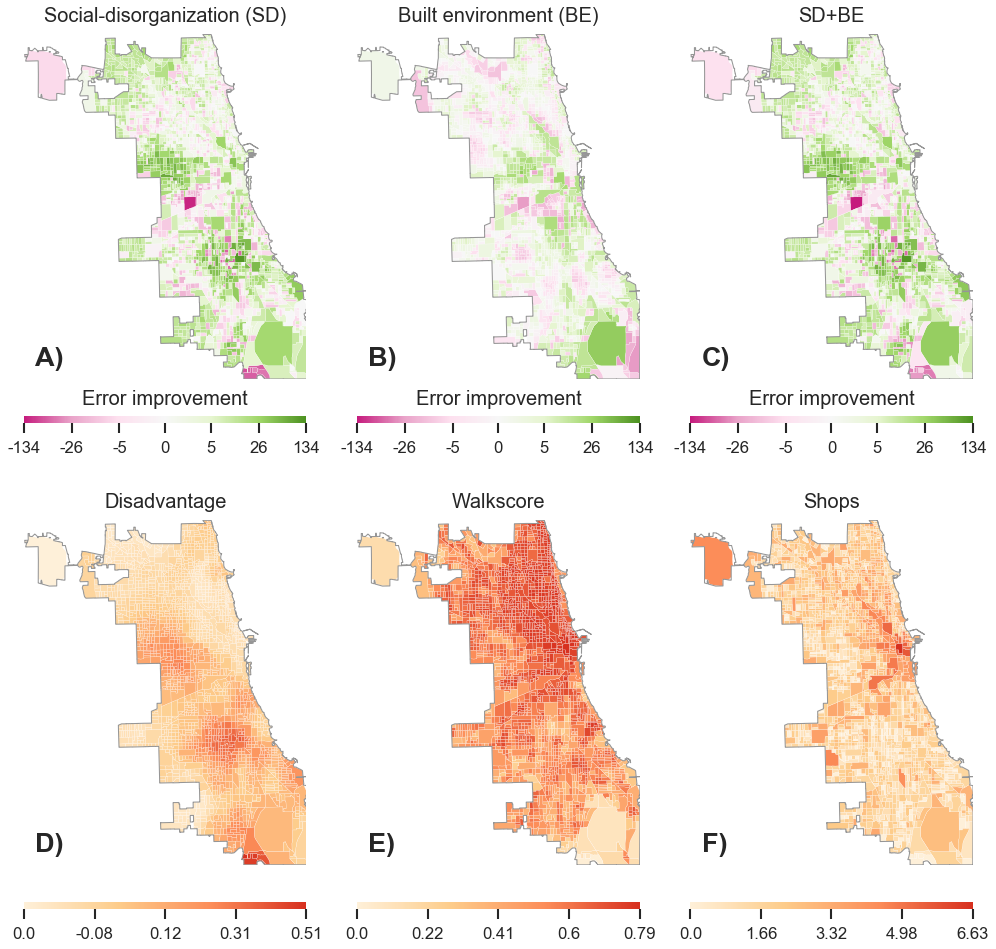}
	\caption{Model improvements in Chicago. Here, we note that we do not possess mobility information, so we compare the SD, BE and SD+BE models. It can be seen that SD model improves the prediction from the core model very consistently, especially in deprived areas, while the BE model seems to improve the prediction in southern Chicago. The SD+BE outperforms the core model and the other models almost everywhere.}
	\label{fig:chicago_improvement}
\end{figure*}

\Cref{fig:boston_improvement}, \Cref{fig:bogota_improvement}, \Cref{fig:LA_improvement} and \Cref{fig:chicago_improvement} show the improvement of each model against the Core model, and some reference variables for each city.

\section{Auto-correlation of features}
\begin{figure}[h!]
	\centering
	\includegraphics[width=0.95\textwidth]{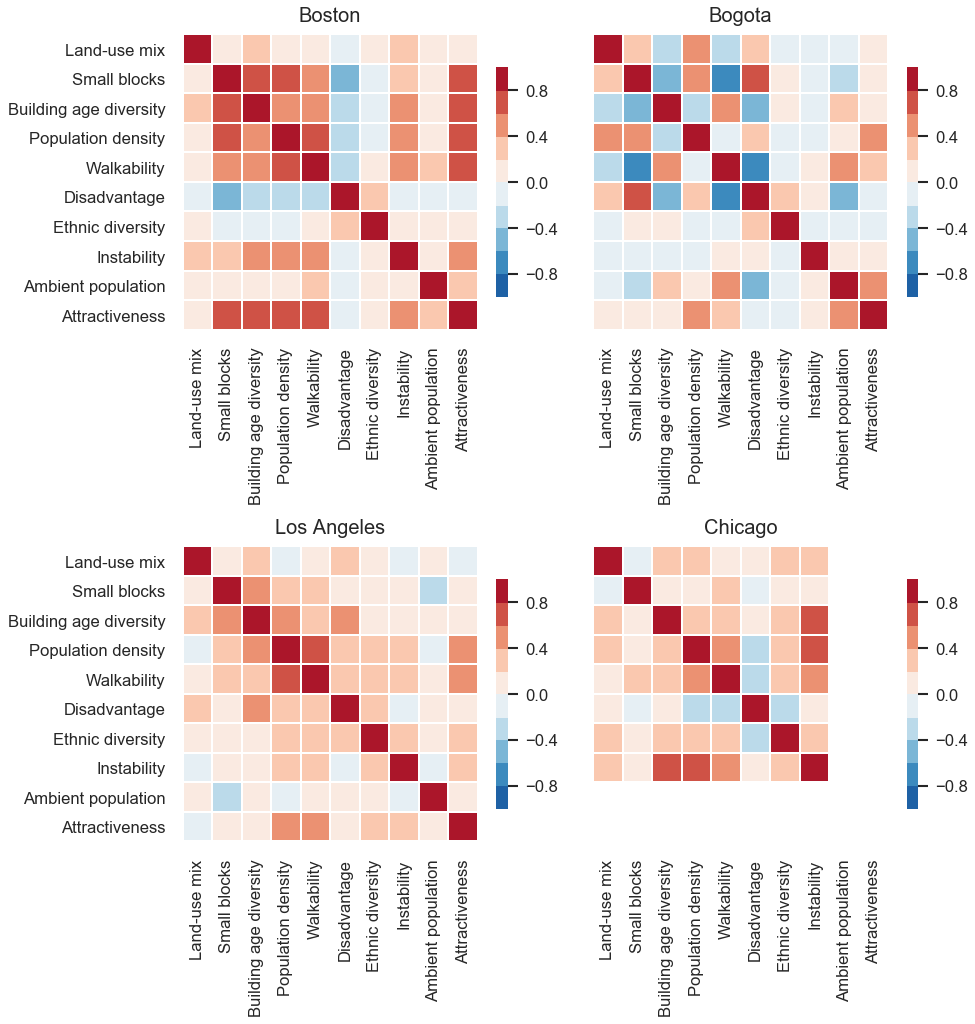}
	\caption{Correlation of features in the different cities.} 
	\label{fig:correlation_features}
\end{figure}

\Cref{fig:correlation_features} shows how the features do not act with the same strength and direction in all cities.

\section{The minimal model}
\Cref{table:minimal} shows the results of the minimal model, which employs only the features that play the same role in all the cities. Results show that no minimal setting is better at predicting crime in all cities.

\begin{table*}[tbhp!]
	\centering
    \footnotesize
	\ra{1.2}
	\begin{tabular}{@{}llrrrrrrrrrrrrrr@{}}
	\toprule
		\textbf{City} & \textbf{Model}  & \textbf{LOO} & $\mathbf{MC_p}$ 
		\\
		\midrule 
		\multirow{2}{*}{Bogota} & Minimal &  -3872 & -0.046\\
		& Full & \textbf{-3809} & -0.040\\
		\midrule
		\multirow{2}{*}{Boston} & Minimal & -1994 & -0.028\\
		& Full \textbf{-1957} & -0.040\\
		\midrule
		\multirow{2}{*}{Los Angeles} & Minimal & -9534 & -0.005\\
		& Full & \textbf{-9453} & 0.003 \\
		\midrule
		\multirow{2}{*}{Chicago} & Minimal & -8009 & 0.012\\
		& Full & \textbf{-8003} & 0.003 \\
    \bottomrule
	\end{tabular}
	\caption{Results of the Full model and the minimal one, which exploits only those features that play the same role in all the cities.}
	\label{table:minimal}
\end{table*}

\section{Corehood tests}
\Cref{table:ego} shows the results of the Full model, and SD+BE model in Chicago, for different sizes of Corehood. From the results we can observe that the best size to infer the neighborhood effect is half a mile.

\begin{table*}[tbhp!]
	\centering
    \footnotesize
	\ra{1.2}
	\begin{tabular}{@{}lccrrrrrrrrrrrrr@{}}
	\toprule
		\textbf{City} & \textbf{Model} & \textbf{Corehood size} & \textbf{LOO} & $\mathbf{MC_p}$ 
		\\
		\midrule
		\multirow{2}{*}{Bogota} & \multirow{2}{*}{Full} & 0.5 miles & \textbf{-3809} & -0.040\\
		& & 1 mile & -3860 & -0.005\\
		\midrule
		\multirow{2}{*}{Boston} & \multirow{2}{*}{Full} & 0.5 miles & \textbf{-1957} & -0.040\\
		& & 1 mile & -2026 & -0.011\\
		\midrule
		\multirow{2}{*}{Los Angeles} & \multirow{2}{*}{Full} & 0.5 miles & \textbf{-9453} & 0.003 \\
		& & 1 mile & -9644 & -0.001 \\
		\midrule
		\multirow{2}{*}{Chicago} & \multirow{2}{*}{SD+BE} & 0.5 miles & \textbf{-8003} & 0.003 \\
		& & 1 mile & -8066 & -0.016\\
    \bottomrule
	\end{tabular}
	\caption{Results of the Full or SD+BE model with different sizes of Corehood.}
	\label{table:ego}
\end{table*}

\end{document}